\documentclass[pre,aps,showpacs,twocolumn,superscriptaddress]{revtex4}
\usepackage{graphicx,dcolumn}
\usepackage{amsfonts,amsmath,amssymb,bm,bbm, mathtools, siunitx}
\usepackage{float}
\usepackage{color}
\usepackage{physics}
\usepackage[toc,page]{appendix}

\def\be{\begin{eqnarray}}
\def\ee{\end{eqnarray}}
\def\ben{\begin{eqnarray*}}
\def\een{\end{eqnarray*}}
\def\bes{\begin{subequations}}
\def\ees{\end{subequations}}
\def\ds{\displaystyle}
\def\nn{\nonumber}

\def\bfq{{\bf q}}
\def\bfk{{\bf k}}
\def\bfkp{{{\bf k}^\prime}}
\newcommand{\wig}[1]{\mathrel{\hbox{\hbox to 0pt{\lower.6ex\hbox{$\sim$}\hss    }\raise.4ex\hbox{$#1$}}}}

\begin{document}

\title{Theory of the electron-ion temperature relaxation rate\\ 
spanning the hot solid metals and plasma phases}

\author{J\'er\^ome \surname{Daligault}}
\email{daligaul@lanl.gov}
\author{Jacopo \surname{Simoni}}
\email{jsimoni@lanl.gov}
\affiliation{Theoretical Division, Los Alamos National Laboratory, Los Alamos, NM 87545, USA}

\begin{abstract}
We present a theory for the rate of energy exchange between electrons and ions -- also known as the electron-ion coupling factor -- in
physical systems ranging from hot solid metals to plasmas, including liquid metals and warm dense matter.
The paper provides the theoretical foundations of a recent work [J. Simoni and J. Daligault, Phys. Rev. Lett. {\bf 122}, 205001 (2019)], where first-principles quantum molecular dynamics calculations based on this theory were presented for representative materials and conditions.
We first derive a general expression for the electron-ion coupling factor that includes self-consistently the quantum mechanical and statistical nature of electrons, the thermal and disorder effects, and the correlations between particles.
The electron-ion coupling is related to the friction coefficients felt by individual ions due to their non-adiabatic interactions with the electrons.
Each coefficient satisfies a Kubo relation given by the time integral of the autocorrelation function of the interaction force of an ion with the electrons.
Exact properties and different representations of the general expressions are discussed.
We then show that our theory reduces to well-known models in limiting cases.
In particular, we show that it simplifies to the standard electron-phonon coupling formula in the limit of hot solids with lattice and electronic temperatures much greater than the Debye temperature, and that it extends the electron-phonon coupling formula beyond the harmonic phonon approximation.
For plasmas, we show that the theory readily reduces to well-know Spitzer formula in the hot plasma limit, to the Fermi golden rule formula in the limit of weak electron-ion interactions, and to other models proposed to go beyond the latter approximation.
We explain that the electron-ion coupling is particularly well adapted to averaged atom models, which offer an effective way to include non-ideal interaction effects to the standard models and at a much reduced computational cost in comparison to first-principles quantum molecular dynamics simulations.
\end{abstract}

\pacs{} 

\date{\today}

\maketitle

\section{Introduction} \label{section_1}

Nonequilibrium states of matter where the constituent electrons and ions are separately strongly driven out of equilibrium are routinely created in the laboratory.
Such conditions typically occur when a material is subjected to an impulsive perturbation such as caused by an intense femtosecond laser pulse \cite{Gamaly2011}, the irradiation by swift neutrons and charged projectiles \cite{Race2010}, or a strong shock wave \cite{Celliers1992}.
Understanding the energy exchanges between and among the electrons and the ions that follow the excitation and drive the system towards equilibirum is a long standing problem in condensed matter \cite{Kaganov1957,Fujimoto1984, Schoenlein1987, Elsayed_et_al_1987, Allen1987, Fann1992, Groeneveld1995, Hohlfeld2000, DelFatti2000,Bonn2000, Rethfeld2002,Lin_et_al_2008,Mueller2013,JiZhang2016, Henighan2016,Waldecker2016,Sadasivam2017,Maldonado2017, Lu1} and plasma physics \cite{Spitzer1962,Ramazashvili1962,Kihara1963,Brysk1975, Hazak2001,DharmaWardana_2001,Gericke2002,Daligault_Mozyrsky_2007,Daligault_Mozyrsky_2008,Glosli2008, DimonteDaligault2008,BrownSingleton2009,DaligaultDimonte2009,Faussurier2010,Vorberger2012,Baalrud2014, Ng2012,Leguay2013,Cho_et_al_2015, Jourdain_et_al_2018, Ogitsu2018, SimoniDaligaultPRL2019}.
Because of the small electron to ion mass ratio, it is customary to distinguish two distinct time scales, namely a short time scale that characterizes the fast internal thermalization of each particle species, and a longer time scale that characterizes the slower equilibration of the electron and ion temperatures.
The latter, which is the subject of this work, is governed by the strength of the electron-ion coupling factor.

For solids and weakly coupled plasmas, the central mechanisms that govern the energy exchanges between electrons and ions have been known for a long time.
In the case of solids, the energy exchanges are well described as resulting from the interactions between electrons and phonons \cite{Allen1987,Patterson2010}.
Yet, the modeling of energy exchanges between electrons and phonons remains an active subject of research driven by increasingly accurate measurements and numerical simulations\cite{Mueller2013,Waldecker2016,Henighan2016,Maldonado2017,Lu1}.
In the case of weakly coupled plasmas, the energy exchanges are well described as resulting from individual binary collisions between charged particles screened by the surrouding plasma \cite{Spitzer1962}.
Different methods have been proposed and validated with simulations to self-consistently take into account the effects of the plasma on binary collisions \cite{Glosli2008,BrownSingleton2009,DimonteDaligault2008,DaligaultDimonte2009,Baalrud2014}.

For systems at the confluence of solids and plasmas, however, 
different models \cite{Lin_et_al_2008,Brysk1975,Hazak2001,DharmaWardana_2001,Daligault_Mozyrsky_2008,Faussurier2010,Vorberger2012} have been proposed that offer diverging predictions even for simple materials (see table I in \cite{SimoniDaligaultPRL2019}).
This intermediate regime, which is characterized by the coexistence and interplay of significant quantum, thermal, disorder and strong Coulomb interaction effects, challenges the standard simplifying approximations of either ordinary condensed-matter physics (e.g., band structure, phonons, etc.) or plasma physics (classical statistics, binary collisions, etc.) \cite{T1}.
The last decade has seen remarkable progress in our ability to form and interrogate in the laboratory materials under conditions at the confluence of solids and plasmas \cite{Fortov2016}.
These experiments typically produce transient, non-equilibrium conditions and measurements may be misleading if recorded while the plasma species are still out of equilibrium.
Remarkably, the electron-ion energy relaxation rate is now accessible to experimental measurements thanks to the diagnostic capabilities offered by the new generation of x-ray light sources \cite{Leguay2013,Cho_et_al_2015,Jourdain_et_al_2018,Ogitsu2018}.

Motivated by these recent developments, we present a theory for the electron-ion coupling factor, the rate of energy exchange between electrons and ions, that applies to physical systems ranging from hot solid metals to plasmas, including liquid metals and warm dense matter.
In a recent paper \cite{SimoniDaligaultPRL2019}, we have presented first-principles calculations based on this theory for representative materials of various electronic complexity and over a range of conditions, but the present theory was only briefly outlined without justification.
The purpose of this paper is to provide a detail exposition of the theory, its properties and its relation to previous models.
We focus on the theoretical aspects only, the numerical algorithms used to obtain the results shown in \cite{SimoniDaligaultPRL2019} will be presented in another publication.
We do not present calculations on specific physical systems in this paper and refer the reader to \cite{SimoniDaligaultPRL2019} and future publications for such applications.

The paper is organized as follows.

In Sec.~\ref{section_2}, we derive the general expression for the electron-ion coupling factor $g$.
The derivation relies on a recent work presented in Ref.~\cite{Daligault_Mozyrsky_2018}, where it was shown, under mild assumptions suitable for the physical systems of interest here, that the coupled dynamics of ions and electrons can be replaced by a simpler, effective classical-quantum dynamics, in which the atomic motions are governed by a stochastic Langevin-like equation and the electron dynamics is described by a master equation for the populations of the electronic states.
By assuming that the electron and ion systems can be characterized by separate temperatures, the theory implies that the two temperatures satisfy coupled rate equations and yields an explicit expression for the relaxation rate -- the electron-ion coupling factor -- in terms of the friction coefficients felt by individual ions due to their non-adiabatic interactions with the electrons.
Each friction coefficient satisfies a Kubo relation given by the time integral of the autocorrelation function of the interaction force of an ion with the electrons.
Several equivalent expressions are given for $g$.

In Sec.~\ref{section_3}, we recast the general result of Sec.~\ref{section_2} in terms of quantities that highlight the many-body screening and correlation effects, and that are more easily amenable to theoretical analysis, practical approximations and numerical evaluations.
We consider three different reformulations, each being based on a different representation of the many-body electron response function in terms of reference response functions, namely the free-particle, the proper and the Kohn-Sham response functions commonly used in condensed matter and plasma physics.
The formulation in terms of the Kohn-Sham response function is at the basis of the first-principles molecular dynamics simulations presented in \cite{SimoniDaligaultPRL2019}, which will be discussed at length elsewhere.

In Sec.~\ref{section_4}, we show that our theory reduces to well-known models in limiting cases and suggests improved practical models.
We show that it reduces to the traditional Spitzer formula in the hot plasma limit \cite{Spitzer1962}, to the Fermi golden rule formula in the limit of weak electron-ion interactions \cite{Hazak2001}, and to the model of Daligault and Dimonte in their attempt to include non-linear electron-ion effects \cite{DaligaultDimonte2009}.
We then explain that the electron-ion coupling is particularly well adapted to averaged atom models, which have proved to be accurate and computationally much more expedient than quantum molecular dynamics simulations.
We then demonstrate that our theory also applies to hot solids, namely to solid metals with lattice and electronic temperatures much greater than the Debye temperature.
The theory reduces to the standard electron-phonon coupling formula \cite{Allen1987} in the appropriate limit and extends the latter by including ionic motions beyond the harmonic approximation.
We finally relate our theory to a simple model due to Wang et al. \cite{Wang1994}, which has served as a reference in recent works on the temperature relaxation rates in hots solids and warm dense matter \cite{Lin_et_al_2008,Cho_et_al_2015}.

For clarity, the technical details are presented in the appendices.
Throughout the paper,  $\hbar$ is the reduced Planck constant, $k_B$ is the Boltzmann constant, and  $m_u$ is the atomic mass unit, and $e^2=q_e^2/4\pi\epsilon_0$, where $q_e$ is the elementary charge and $\epsilon_0$ is the vacuum permittivity.
$\mathrm{Re}z$ and $\mathrm{Im}z$ denote the real and imaginar parts of a complex number $z$.
Throughout the paper, $t$ and $\omega$ denote the time and frequency variables and, with no risk of confusion, $f(\omega)=\int_{-\infty}^\infty{dt e^{i\omega t} f(t)}$ denotes the Fourier transform of the function $f(t)$.
Finally, 
\be
v_{\rm C}({\bf r})=\frac{e^2}{|{\bf r}|}
\ee
denotes the Coulomb potential energy.

\section{Derivation and general expressions of the electron-ion coupling factor} \label{section_2}

\subsection{Definitions and assumptions}

We consider a material containing one atomic species enclosed in a three-dimensional cubic box of volume $V=L^3$.
The material is described as a two-component system comprised of ions (mass $m_i=Am_u$, number density $n_i=N/V$, charge $Ze$) and of electrons (mass $m_e$, density $n_e=Zn_i$), where each ion consists of an atomic nucleus and its most tightly bound, unresponsive core electrons.
Throughout the paper, the calculations are performed by imposing three-dimensional periodic (Born-von Karman) boundary conditions along each direction of the box; physical results are then obtained in the thermodynamic limit where both $N$ and $V$ tend to infinity in such a way that $n_i$ remains constant.

The total Hamiltonian of the system is
\be
{H}&=&\sum_{I=1}^{N}{\frac{{\bf P}_I^2}{2M}}+\underbrace{\sum_{I,J\ne I=1}^N{v_{ii}({\bf R}_I-{\bf R}_J)}}_{\ds\equiv V_{ii}}+{H}_\mathrm{e}({\bf R}) \label{complete_H}
\ee
with the electron Hamiltonian
\ben
{H}_\mathrm{e}({\bf R})&=&\sum_{i=1}^{N_e}{\frac{{\bf p}_i^2}{2m_e}}\\
&&\quad+\underbrace{\sum_{i=1}^{N_e}{\sum_{I=1}^{N}{v_{ie}({{\bf r}}_i-{\bf R}_I)}}}_{\ds\equiv V_{ie}}+\underbrace{\sum_{i,j\ne i}^{N_e}{v_C(|{{\bf r}}_i-{{\bf r}}_j|)}}_{\ds\equiv V_{ee}}\,,
jacop\een
where $v_{ii}$ and $v_{ie}$ are the ion-ion and ion-electron interaction energies.
For simplicity of exposition, the electron-ion interaction is described by a local pseudopotential $v_{ie}(r)$; in practice, the formalism allows to deal with more elaborate descriptions, e.g., using plane-augmented wave pseudopotentials as will be discussed elsewhere.

Below, $R_{\alpha=Ix}$ denotes the position of ion $I$ along the $x$-direction and ${\bf R}=\{R_\alpha\}_{\alpha=1,\dots,3N}$ denotes the set of all ionic positions.

\subsection{Effective dynamics of ions and electrons}\label{section_2_1}

For the metallic systems of interest here, the dynamics governed by the complete quantum Hamiltonian (\ref{complete_H}) can be replaced by a simpler, effective classical-quantum dynamics by making use of the naturally small electron-ion mass ratio and the existence of a manifold of infinitesimally separated electronic excitations.
More precisely, we make the following three assumptions:\\
\noindent\hspace*{0.25cm} (i) The dynamics of each ion can be described by that of the center ${\bf R}_i(t)$ of its narrowly localized wavepacket.
This is justified here, since the thermal de Broglie wavelength $\Lambda=\hbar\sqrt{2\pi/m_ik_BT_i}(\simeq 0.3/\sqrt{AT_{i}[eV]}$ Bohr) of ions is generally much smaller than the spatial variations of forces acting on them due to their large mass and the relatively high temperatures.\\
\hspace*{0.25cm} (ii) The typical ion velocities are small compared to the typical electronic velocities.
For instance, for the two-temperature systems considered later, we assume $T_i/m_i\!\ll\! T_F/m_e$ or $T_i/m_i\! \ll\! T_e/m_e$ in the degenerate $T_e/T_F\!\ll\! 1$ or non-degenerate limit $T_e/T_F\!\gg\! 1$, respectively, where $T_F=\frac{\hbar^2}{2m_e k_B}(3\pi^2 n_e)^{\frac{2}{3}}$ ($\simeq 1.69\,\left(n_e[{\rm cm}^{-3}]/10^{22}\right)^{\frac{2}{3}}$ eV) is the electronic Fermi temperature.
This condition is generally respected due to the natural smallness of $m_e/m_i$, and is challenged only if $T_i\gg T_e$.\\
\hspace*{0.25cm} (iii) We assume that there is a quasi-continuum of electronic states, as it is the case for the metallic systems of interest here.

Under these conditions, the electron and ion dynamics can be described by the coupled set of equations \cite{Daligault_Mozyrsky_2018}
\bes
\be
\ds\frac{d P_n}{dt}&=&\sum_m ( W_{nm} P_m - W_{mn} P_n ) \label{Eq:1b}\\
\ds M\ddot{R}_\alpha&=&-\frac{\partial V_\mathrm{ii}}{\partial R_\alpha} + F_\alpha^\mathrm{BO} - M\sum_{\beta=1}^{3N}\gamma_{\alpha\beta}^{[\mathbf{R}]}\dot{R}_\beta + \xi_\alpha^{[\mathbf{R}]} \,.\quad\quad\quad\label{Eq:1}
\ee
\label{globalcoupleddynamics}
\ees

The electron dynamics is described by the master equation (\ref{Eq:1b}) for the populations $P_n(t)$ of the adiabatic electronic states $\ket{n(\mathbf{R}(t))}$ defined by $\hat{H}_\mathrm{e}(\mathbf{R}(t))\ket{n(\mathbf{R}(t))} = E_n(\mathbf{R}(t))\ket{n(\mathbf{R}(t))}$.
The transition rates between different electronic states are given by
\begin{equation}
  W_{nm} = 2\pi\hbar|\mathbf{d}_{nm}\cdot\mathbf{V}|^2 e^{-\frac{(E_{m}-E_n)|\mathbf{d}_{nm}|^2}{2M|\mathbf{d}_{nm}\cdot\mathbf{V}|^2}}\delta(E_n-E_m),
\end{equation}
where $\mathbf{V}=\dot{\mathbf{R}}$ represents the full set of atomic velocities and the non adiabatic couplings $\mathbf{d}_{nm}=\mel{n}{\nabla_\mathbf{R}}{m}$.

Each ionic position follows a stochastic Langevin-like equation (\ref{Eq:1}), where
\ben
F_\alpha^\mathrm{BO}(t) = \sum_n P_n(t) f_{nn}^\alpha(t),
\een
is the adiabatic Born-Oppenheimer force, which includes the interactions between ions and with the instantaneous electrostatic potential of electrons, where $f_{nn}^\alpha(t)=\mel{n}{-\nabla_{R_\alpha}\hat{H}_\mathrm{e}({\bf R}(t))}{n}$ represents the force exerted on the degree of freedom $R_\alpha$ by the state $\ket{n(\{\mathbf{R}(t)\})}$.
The other terms describe the effect of non-adiabatic transitions between closely spaced electronic states induced by the atomic motions and electronic excitations.
These terms, which are not accounted for in current quantum molecular dynamics simulations, are responsible for the constant, non-reversible, energy exchanges between electron and ions.
Like the buffeting of light liquid particles on a heavy Brownian particle, the non-adiabatic effects produce the friction forces $- M\gamma_{\alpha\beta}^{[\mathbf{R}]}\dot{R}_\beta$, where 
\begin{equation}\label{Eq:1c}
  \gamma_{\alpha\beta}^{[\mathbf{R}]} = -\frac{\pi\hbar}{M} \sum_{n\neq m}\frac{P_n - P_m}{E_n-E_m}f_{nm}^\alpha f_{mn}^\beta\delta(E_n-E_m),
\end{equation}
Here, the out-of-diagonal force matrix elements are defined as $f_{nm}^\alpha = \mel{n}{-\nabla_{R_\alpha}\hat{H}_\mathrm{e}({\bf R}(t))}{m}$.
The symbol $[\mathbf{R}]$ is used to indicate that the quantity depends on the instantaneous atomic configuration of the system. 
However, in order to avoid cluttering the mathematical expressions, we do not always indicate the dependence on $[\mathbf{R}]$.

The second term is a $\delta$-correlated Gaussian random fluctuating force, $\xi_\alpha^{[\mathbf{R}]}(t)$, that satisfies the following two conditions
\begin{eqnarray}
  \ll\xi_\alpha^{[\mathbf{R}]}(t)\gg & = & 0,\\
  \ll\xi_\alpha^{[\mathbf{R}]}(t)\xi_\beta^{[\mathbf{R}]}(t')\gg & = & \frac{B_{\alpha\beta}^{[\mathbf{R}]}}{2}\delta(t - t'),\label{Eq:1d}
\end{eqnarray}
where $\ll\ldots\gg$ denotes an average over the Gaussian noise, and
\begin{equation}\label{Eq:1e}
  B_{\alpha\beta}^{[\mathbf{R}]} = \pi\hbar\sum_{n\neq m}(P_n + P_m)f_{nm}^\alpha f_{mn}^\beta\delta(E_n-E_m)\,.
\end{equation}

\subsection{Application to a two-temperature plasma}

\subsubsection{General result}

For our present purpose, we assume throughout the rest of the paper that the material can be described as an isolated, homogeneous, two-temperature system characterized at all times $t$ by the temperatures $T_e(t)$ and $T_i(t)$ of the electronic (e) and ionic (i) subsystems.
(We could in principle relax the homogeneity condition and add an external energy source, e.g. a laser, but this is not necessary here.)
The instantaneous electronic populations are then $P_n=e^{-E_n/k_BT_e }/\mathcal{Z}$, with the canonical partition function $\mathcal{Z}=\mathrm{Tr}e^{-\hat{H}_\mathrm{e}^{[\mathbf{R}]}/k_BT_e}=\sum_n{e^{-E_n/k_BT_e}}$. 

The temporal evolution of the temperatures can be obtained by applying the evolution equations (\ref{Eq:1})-(\ref{Eq:1b}) to the ensemble averaged kinetic energy of the ions $K_\mathrm{ion}(t)= \left\langle\frac{1}{2}M\dot{\bf R}^2(t)\right\rangle$  and to the internal energy of electrons $E_\mathrm{elec}(t)= \Big\langle\sum_n{P_n(t)E_n(t)}\Big\rangle$.
As shown below, this yields the rate equations
\bes
\be 
\ds c_\mathrm{i}^{0}\frac{dT_\mathrm{i}}{dt}&=&g[T_\mathrm{e}(t) - T_\mathrm{i}(t)] \label{Eq:1f}\\
\ds c_\mathrm{e}\frac{dT_\mathrm{e}}{dt}&=& - g[T_\mathrm{e}(t) - T_\mathrm{i}(t)] \,,\label{Eq:1g}
\ee
\label{temperaturerateequations}
\ees
\noindent where $c_\mathrm{i}^0=V^{-1}\partial [3Nk_BT_i/2]/\partial T_i=3n_ik_B/2$ is the kinetic component of the heat capacity of the ions, $c_\mathrm{e}=V^{-1}\partial \big\langle \sum_n{P_nE_n}\big\rangle /\partial T_e$ is the electronic heat per unit volume, and $g$ is the electron-ion coupling factor of interest here given by
\begin{equation}\label{Eq:2f}
  g(T_\mathrm{e},T_\mathrm{i})= 3 k_\mathrm{B}n_\mathrm{i}\bigg\langle\frac{1}{3N}\sum_{\alpha=1}^{3N}\gamma_{\alpha\alpha}^{[\mathbf{R}]}(T_e,T_i)\bigg\rangle\,.
\end{equation}

We show in Sec.~\ref{Kubo_relations} that the friction coefficients $\gamma_{\alpha\alpha}^{[\mathbf{R}]}$ and the resulting coupling factor $g$ can be written in the form of standard Kubo relations like the ordinary electronic and ionic transport coefficients.
The physical content of Eq.(\ref{Eq:2f}) is discussed in the following sections.
Before, we present a derivation of Eqs.(\ref{temperaturerateequations}) and (\ref{Eq:2f}).

\subsubsection{Proof of Eqs.(\ref{temperaturerateequations}) and (\ref{Eq:2f}).}

Equation (\ref{Eq:1f}) for the ionic temperature can be readily obtained by recalling the close relationship between the Langevin equation and the Fokker-Planck-Kramers equation \cite{vanKampenbook}.
The swarm of trajectories generated by Eq.(\ref{Eq:1}) can be described by the probability distribution function 
\be
f({\bf R},{\bf V},t)=\left\langle \delta\Big({\bf R}-{\bf R}(t)\Big)\delta\Big({\bf V}-\dot{{\bf R}}(t)\Big) \right\rangle\,,
\ee
with ${\bf R},{\bf V}\in \mathbb{R}^{3N}$, and $\langle\dots\rangle$ represents the double average $\langle\ll\dots\gg\rangle$ over the noise and over the initial distribution function.
The distribution function $f$ satisfies the Fokker-Planck-Kramers equation
\be
\lefteqn{\frac{\partial f}{\partial t}+\sum_{\alpha=1}^{3N}{V_\alpha\frac{\partial f}{\partial R_\alpha}}+\sum_{\alpha=1}^{3N}\frac{F_\alpha^{BO}}{M}\frac{\partial f}{\partial V_\alpha}}&& \label{FokkerPlanckKramersequation}\\
&&=\sum_{\alpha,\beta=1}^{3N}{\frac{\partial}{\partial V_\alpha}\left[\gamma_{\alpha\beta}\left(V_{\beta}f\right)+\frac{B_{\alpha,\beta}}{2M^2}\frac{\partial}{\partial V_{\beta}}f\right]}\,.\nn
\ee
Remark that this equation, which governs the full distribution functions of the ions, should not be confused with the celebrated Fokker-Planck equation for the single-particle distribution function that is widely used in ordinary plasma physics \cite{ThomsonHubbard1960}.
From this evolution equation, we find the time evolution of the kinetic energy of ions $K_{ion}(t)=\iint{\frac{1}{2}M{\bf V}^2 f({\bf R},{\bf V},t) d{\bf R}d{\bf V}}$,
\be
\frac{dK_{ion}}{dt}&=&\sum_{\alpha=1}^{3N}\iint{V_\alpha F_\alpha^{BO} f({\bf R},{\bf V},t) d{\bf R}d{\bf V}}\label{dKiondt}\\
&+&\sum_{\alpha=1}^{3N}\iint{\left[-\gamma_{\alpha\alpha}MV_\alpha^2+\frac{B_{\alpha\alpha}}{2M}\right]f({\bf R},{\bf V},t) d{\bf R}d{\bf V}}\,.\nn
\ee
By assuming a Maxwellian velocity distribution at temperature $T_i(t)$, i.e. $f({\bf R},{\bf V},t)\propto e^{-M{\bf V}^2/2k_BT_i(t)}$, the kinetic energy $K_{ion}(t)=3Nk_BT_i(t)/2$ and Eq.(\ref{dKiondt}) simplifies
\be
\frac{dK_{ion}}{dt}&=&\frac{3Nk_B}{2}\frac{dT_i(t)}{dt}\label{dTidt_preresult}\\
&=&-\bigg\langle\sum_{\alpha=1}^{3N}\gamma_{\alpha\alpha}\bigg\rangle k_B T_i+\bigg\langle\sum_{\alpha=1}^{3N}\frac{B_{\alpha\alpha}}{2M}\bigg\rangle\,. \nn
\ee
Since we also assume a thermal distribution of adiabatic states at temperature $T_e$, $\gamma_{\alpha\alpha}$ and $B_{\alpha\alpha}$ satisfy the relation $B_{\alpha\alpha}=2M k_BT_e\gamma_{\alpha\alpha}$ (compare Eqs.(\ref{Eq:1c}) and (\ref{Eq:1e}), see details in \cite{Daligault_Mozyrsky_2018}), and Eq.(\ref{dTidt_preresult}) reduces to the desired result, Eq.(\ref{Eq:1f}).

The equation (\ref{Eq:1g}) for the electronic temperature is obtained by combining Eq.(\ref{dTidt_preresult}) with the conservation equation $\frac{d}{dt}\left\langle \frac{1}{2}M{\dot{\bf R}}^2+V_{ii}+\sum_n{P_nE_n}\right\rangle=0$ and with the property $d \langle V_{ii}\rangle/dt=0$ that is easily shown using the Fokker-Planck-Kramers equation.
A proof of the energy conservation equation can be found in Ref.~\cite{Daligault_Mozyrsky_2018}.
This yields the desired result for the rate of change of the electronic energy $E_\mathrm{elec}$,
\be
\frac{1}{V}\frac{d}{dt}\langle E_\mathrm{elec}\rangle=-g(T_e-T_i)\,,
\ee
or, with $c_\mathrm{e}\equiv V^{-1}\partial \langle E_\mathrm{elec}\rangle/\partial T_e$, 
\be
c_e\frac{dT_e}{dt}=-g(T_e-T_i)\,.
\ee 

\subsection{Kubo relations for the friction coefficients and the electron-ion coupling. Sum rule.} \label{Kubo_relations}

Under the two temperature assumption considered here, the electronic populations are
\be
P_n=\frac{e^{-E_n/k_BT_e }}{\mathcal{Z}}\,, \label{thermal_Pn}
\ee
and the general expression (\ref{Eq:1c}) for the friction coefficients can be effectively written compactly in the form of ordinary Kubo relations, i.e. as time integrals of correlation functions.
The different expressions below result from well-known relations between thermal correlation functions, response function and their Lehmann representations \cite{Kubo_book}; their defintions are recalled for convenience in appendix~\ref{appendix_1}.
Below, $\langle\ldots\rangle_\mathrm{e}=\mathrm{Tr}\bigg[\frac{e^{-\hat{H}_\mathrm{e}/k_BT_e}}{\mathcal{Z}}\ldots\bigg]$ indicates the thermal average over the electronic subsystem.\\

\paragraph*{i)} With Eq.(\ref{thermal_Pn}), the expression (\ref{Eq:1c}) can be written compactly as (see Eq.(\ref{K_lehmann}))
\begin{equation} \gamma_{\alpha\beta}^{[\mathbf{R}]}=\frac{1}{2Mk_BT_e}\int_{-\infty}^\infty{K_{\alpha\beta}^{[\mathbf{R}]}(t)\,dt}=\frac{1}{2Mk_BT_e}K_{\alpha\beta}^{[\mathbf{R}]}(\omega=0) \label{gammaab_Kubocorrelation}
\end{equation}
in terms of the Kubo correlation function
\begin{equation}
  K_{\alpha\beta}^{[\mathbf{R}]}(t)=k_BT_e\int_0^{1/k_BT_e} d\lambda \left\langle\, e^{\lambda\hat{H}_\mathrm{e}^{[\mathbf{R}]}}\delta\hat{f}_\beta e^{-\lambda\hat{H}_\mathrm{e}^{[\mathbf{R}]}}\delta\hat{f}_\alpha(t)\,\right\rangle_\mathrm{e},
\end{equation}
where $\hat{f}_{\alpha=Ix}(t)=-e^{i\hat{H}_\mathrm{e}t/\hbar}\,\partial_{R_\alpha}\hat{H}_\mathrm{e}\, e^{-i\hat{H}_\mathrm{e}t/\hbar}$ is the force operator at time $t$ between ion $I$ and the electronic subsystem along the $x$-direction and $\delta\hat{f}_\alpha(t)$ indicates the same operator deprived of its diagonal matrix elements, i.e. $(\delta f_{\alpha})_{nm}=f_{nm}^{\alpha}(1-\delta_{n,m})$.\\

\paragraph*{ii)} Using the property (\ref{K_C_relation}), the relation (\ref{gammaab_Kubocorrelation}) can be written in terms of the electron-ion force-force correlation function $\langle\,\delta\!\hat{f}_\alpha(t)\delta\!\hat{f}_\beta(0)\,\rangle_\mathrm{e}$, 
\begin{equation}
  \gamma_{\alpha\beta}^{[\mathbf{R}]} = \frac{1}{2Mk_BT_e}\mathrm{Re}\int_0^\infty dt\langle\,\delta\!\hat{f}_\alpha(t)\delta\!\hat{f}_\beta(0)\,\rangle_\mathrm{e}\,.\label{gammaab_correlation}
\end{equation}

\paragraph*{iii)} Equation (\ref{gammaab_correlation}) can be expressed in terms of the symmetric electronic density correlation function $S_\mathrm{ee}^{[\mathbf{R}]}(\mathbf{r}_1,\mathbf{r}_2,t)=\frac{1}{2}\langle\,\delta\hat{n}_\mathrm{e}(\mathbf{r}_1,t)\delta\hat{n}_\mathrm{e}(\mathbf{r}_2,0)+\delta\hat{n}_\mathrm{e}(\mathbf{r}_2,0) \delta\hat{n}_\mathrm{e}(\mathbf{r}_1,t)\,\rangle_\mathrm{e}$ as follows
\be 
\lefteqn{\gamma_{\alpha\beta}^{[\mathbf{R}]} = \frac{1}{Mk_BT_e}}&& \label{gammaalphabetaSee}\\
&&\times\int_V d{{\bf r}_1}\int_V d{{\bf r}_2}f_\alpha(\mathbf{r}_1)S_\mathrm{ee}^{[\mathbf{R}]}(\mathbf{r}_1,\mathbf{r}_2,\omega=0)f_\beta(\mathbf{r}_2)\,.\nn
\ee 
where
\be
f_{\alpha=Ix}(\mathbf{r}) = -\nabla_{R_{\alpha=Ix}}v_{ie}(\mathbf{r}-{\bf R}_I)\,,
\ee
is the force along the $x$-direction between the ion $I$ and an electron located at ${\bf r}$.
Equation (\ref{gammaalphabetaSee}) is easily obtained using $\partial_{{\bf R}_I}\hat{H}_\mathrm{e}=\int_V{d{\bf r}\,\partial_{{\bf R}_I}\,v_{ie} (\mathbf{r}-{\bf R}_I)\hat{n}_\mathrm{e}(\mathbf{r})}$ in Eq.(\ref{gammaab_correlation}).\\

\paragraph*{iv)} The expression (\ref{gammaab_Kubocorrelation}) can also be written in terms of the electron response function (a.k.a. susceptibility), $\chi_\mathrm{ee}^{[\mathbf{R}]}(\mathbf{r}_1,\mathbf{r}_2,t)=-\frac{i}{\hbar}\theta(t)\langle\,[\delta\hat{n}_\mathrm{e}(\mathbf{r}_1,t),\delta\hat{n}_\mathrm{e}(\mathbf{r}_2,0)]\,\rangle_\mathrm{e}$ of the electronic subsystem in the frozen ionic configuration ${\bf R}$, as follows
\be
\lefteqn{\gamma_{\alpha\beta}^{[\mathbf{R}]} =-\frac{1}{M}}&&\label{Eq:5}\\
&&\times\int_V d{{\bf r}_1}\int_V d{{\bf r}_2}f_\alpha(\mathbf{r}_1)\partial_\omega\mathrm{Im}\chi_\mathrm{ee}^{[\mathbf{R}]}(\mathbf{r}_1,\mathbf{r}_2,\omega=0)f_\beta(\mathbf{r}_2)\,.\nn
\ee
This is easily found using Eq.(\ref{gammaalphabetaSee}) and the fluctuation-dissipation relation (\ref{SChifluctuationdissipation}) between $\chi_\mathrm{ee}^{[\mathbf{R}]}$ and $S_\mathrm{ee}^{[\mathbf{R}]}$.
We shall mainly rely on the expressions (\ref{gammaab_correlation}) and  (\ref{Eq:5}) in the reminder of the paper.\\

\paragraph*{v)} With the help of the previous expressions, the electron-ion coupling factor (\ref{Eq:2f}) writes as
\begin{widetext}
\bes
\be
g(T_e,T_i)&=&\frac{3n_\mathrm{i}}{2MT_e}\left\langle\frac{1}{3N}\sum_{\alpha=1}^{3N}\mathrm{Re}\int_0^\infty dt\left\langle\,\delta\hat{f}_\alpha(t)\delta\hat{f}_\alpha(0)\,\right\rangle_\mathrm{e}\right\rangle\\
&=&-\frac{3 k_\mathrm{B}n_\mathrm{i}}{M}\left\langle\frac{1}{3N}\sum_{\alpha=1}^{3N}
\int_V d{{\bf r}_1}\int_V d{{\bf r}_2}f_\alpha(\mathbf{r}_1)\partial_\omega\mathrm{Im}\chi_\mathrm{ee}^{[\mathbf{R}]}(\mathbf{r}_1,\mathbf{r}_2,\omega=0)f_\alpha(\mathbf{r}_2)
\right\rangle, \label{g_Kubo_b}
\ee
\label{g_Kubo}
\ees
\end{widetext}

\paragraph*{vi)} The electron-ion coupling (\ref{Eq:2f}) equals the trace of the matrix $\tensor{\gamma}^{[\mathbf{R}]}=\left\{\gamma^{[\mathbf{R}]}_{\alpha,\beta}\right\}$ of friction coefficients.
Other combinations of matrix elements satisfy remarkable properties.
Most remarkably,
\begin{equation}\label{Eq:9}
 \sum_\mathrm{I,J}\gamma^{[\mathbf{R}]}_{\mathrm{I}x,\mathrm{J}y}=0\quad\text{for all }x,y\,,
\end{equation}
and, therefore, 
\be
\sum_{\alpha,\beta}\gamma^{[\mathbf{R}]}_{\alpha\beta}=0\,. \label{Eq:9p}
\ee
for the sum of all matrix elements.
These sum rules, which are physically related to the conservation of momentum, are proved in appendix~\ref{appendix_3}.

\section{Electronic screening, exchange and correlation effects} \label{section_3}

Thus far we have given general expressions for the friction coefficients and the electron-ion coupling factor in which the electrons are not described individually but are described as a single entity.
For instance, the electronic states $|n({\bf R})\rangle$ in Eq.(\ref{Eq:1c}) are many-body states, and the response function $\chi_\mathrm{ee}^{[\mathbf{R}]}$ in Eq.(\ref{Eq:5}) is the full density-density response function of the electronic subsystem.
In this section, we recast these results in terms of quantities that instead emphasize the individual character of electrons.
The many-body screening and correlation effects are displayed more distinctly in terms of dielectric functions and local field corrections.
To accomplish this, we express the full density-density response function $\chi_\mathrm{ee}^{[\mathbf{R}]}$ in terms of the response of a reference system of independent particles.
We shall consider three different reference response functions that are often used \cite{GiulianiVignale2005}, namely the proper response function, the non-interacting response function, and the Kohn-Sham response function.
As we shall show in Sec.~\ref{section_4}, previous models for $g$ are easily recovered from these reformulations.

For pedagogical clarity, we follow the same line of presentation in each case.
To this end, we recall that the response function $\chi_\mathrm{ee}^{[\mathbf{R}]}$ gives the change in the ground-state electronic density $n_e^{(0)}$ through
\be
\delta n_e({\bf r},\omega)&=&n_e({\bf r},\omega)-n_e^{(0)}({\bf r},\omega)\nn\\
&=&\int_V{d{\bf r}^\prime \chi_{\rm ee}^{[\mathbf{R}]} ({\bf r},{\bf r}';\omega)\delta v_{\rm ext}({\bf r}',\omega)}\,, \label{delta_ne}
\ee
when the electron subsystem in the frozen ionic configuration ${\bf R}$ is perturbed by a weak time-dependent scalar potential $\delta v_{ext}({\bf r},t)$ \cite{GiulianiVignale2005}.

\subsection{Relation to the proper response function}

The proper density-density response function $\tilde{\chi}^{[\mathbf{R}]}$ allows to write Eq. (\ref{delta_ne}) as \cite{GiulianiVignale2005}
\begin{equation}
\delta n_e({\bf r},\omega)=\int_V{d{\bf r}^\prime \tilde{\chi}^{[\mathbf{R}]} ({\bf r},{\bf r}';\omega)\delta v_{\rm sc}({\bf r}',\omega)}
\end{equation}
in terms of the screened potential
\begin{equation} \label{v_sc}
\delta v_{\rm sc}({\bf r},\omega)=\delta v_{\rm ext}({\bf r},\omega) +\int_V{d{\bf r}' v_{\rm C}({\bf r}-{\bf r}') \delta n_e({\bf r}',\omega)}.
\end{equation}
This is the potential experienced by a test particle (i.e., a fictitious particle that does not disturb the system in which it is embedded) in the electron gas due to both the external potential and the Coulomb field created by the density pertubation $\delta n_e$ induced by $\delta v_{ext}$.
This potential does not account for the correlation that exists between a given electron of the electron gas and the other electrons.
By definition, these correlation effects are implicitely incorporated in the proper response function $\tilde{\chi}^{[\mathbf{R}]}$.

The proper response function $\tilde{\chi}^{[\mathbf{R}]}$ is related to the full response function $\chi_{\rm ee}^{[\mathbf{R}]}$ through the integral (Dyson) equation,
\be
\lefteqn{\chi_\mathrm{ee}^{[\mathbf{R}]}(\mathbf{r}_1,\mathbf{r}_2,\omega) = \tilde{\chi}^{[\mathbf{R}]}(\mathbf{r}_1,\mathbf{r}_2,\omega)}&&  \label{Dysonproper}\\
&&+ \int_V d{{\bf r}}\int_V d{{\bf r}'}\tilde{\chi}^{[\mathbf{R}]}(\mathbf{r}_1,\mathbf{r},\omega) v_\mathrm{C}(\mathbf{r}-\mathbf{r'})\chi_\mathrm{ee}^{[\mathbf{R}]}(\mathbf{r'},\mathbf{r}_2,\omega)\,.\nn
\ee
Using this relation in Eq.(\ref{Eq:5}), it is straigthforward to write the friction coefficient $\gamma_{\alpha\beta}^{[\mathbf{R}]}$ as (see appendix~\ref{appendix_2p}) 
\be 
\lefteqn{\gamma_{\alpha\beta}^{[\mathbf{R}]} =-\frac{1}{M}}&& \label{Eq:5b}\\
&\times&\int_V d{{\bf r}_1}\int_V d{{\bf r}_2}f_\alpha^\mathrm{L}(\mathbf{r}_1)\partial_\omega \mathrm{Im}\tilde{\chi}^{[\mathbf{R}]}(\mathbf{r}_1,\mathbf{r}_2,\omega=0)f_\beta^\mathrm{R}(\mathbf{r}_2)\,, \nn
\ee
%
where the $f_{\alpha=Ix}^\mathrm{L(R)}$ represents the force of interaction along the $x$-direction of ion $I$ and a test charge,
\begin{eqnarray}
  f_\alpha^\mathrm{L}(\mathbf{r}_1) & = & \int_V d{{\bf r}} f_\alpha(\mathbf{r})\varepsilon_\mathrm{L}^{[\mathbf{R}]}(\mathbf{r},\mathbf{r}_1,\omega=0)^{-1},\\
  f_\alpha^\mathrm{R}(\mathbf{r}_1) & = & \int_V d{{\bf r}} \varepsilon_\mathrm{R}^{[\mathbf{R}]}(\mathbf{r}_1,\mathbf{r},\omega=0)^{-1}f_\alpha(\mathbf{r}),
\end{eqnarray}
in terms of the inverse of the left (L) and right (R) dielectric functions
\be 
 \varepsilon_\mathrm{L}^{[\mathbf{R}]}(\mathbf{r},\mathbf{r}_1,\omega)=\delta({\mathbf{r}-\mathbf{r}_1}) - \int_V d{{\bf r}'}\tilde{\chi}^{[\mathbf{R}]}(\mathbf{r},\mathbf{r'},\omega)v_\mathrm{C}(\mathbf{r'}-\mathbf{r}_1)&\nn\\
  \varepsilon_\mathrm{R}^{[\mathbf{R}]}(\mathbf{r},\mathbf{r}_1,\omega)=\delta({\mathbf{r}-\mathbf{r}_1}) - \int_V d{{\bf r}'}v_\mathrm{C}(\mathbf{r}-\mathbf{r'})\tilde{\chi}^{[\mathbf{R}]}(\mathbf{r'},\mathbf{r}_1,\omega)&\nn
\ee 
Note that the definition of the left (L) and right (R) dielectric functions is needed at this level of generality since the system is embedded in the inhomogenous background of the ionic configuration ${\bf R}$ (for homogeneous systems, $\epsilon _\mathrm{L} =\epsilon _\mathrm{R}$ (see Sec.~\ref{section_4})).

With Eq.(\ref{Eq:5b}), the electron-ion coupling factor becomes
\begin{widetext}
\begin{eqnarray}
g(T_e,T_i)&=&-\frac{3 k_\mathrm{B}n_\mathrm{i}}{M}\left\langle\frac{1}{3N}\sum_{\alpha=1}^{3N}
\iint_V\! d{{\bf r}_1}d{{\bf r}_2}f_\alpha^\mathrm{L}(\mathbf{r}_1)\partial_\omega \mathrm{Im}\tilde{\chi}^{[\mathbf{R}]}(\mathbf{r}_1,\mathbf{r}_2,\omega=0)f_\alpha^\mathrm{R}(\mathbf{r}_2)
\right\rangle\,. \label{g_proper}
\end{eqnarray}
\end{widetext}
The expressions (\ref{Eq:5b}) and (\ref{g_proper}) make a good starting point of further theoretical analysis because the proper response function -- or irreducible reponse function -- lends itself well to advanced perturbative methods in order to systematically include the correlation effects beyond the mean-field approximation \cite{GiulianiVignale2005,Mahan2000,BruusFlensberg2004}.
This goes beyond the scope of the present work and, instead, we shall now consider another representation where these important many-body effects appear even more visibly in terms of local field corrections.

\subsection{Relation to the ideal gas response function} \label{section_3_b}

It is also common to express the deviation $\delta n_e$ as the induced density of a noninteracting (free) electron gas such as \cite{GiulianiVignale2005}
\begin{equation}
\delta n_e({\bf r},\omega)=\int{d{\bf r}^\prime \chi_0^{[\mathbf{R}]} ({\bf r},{\bf r}';\omega)\delta v_{\rm eff}({\bf r}',\omega)}
\end{equation}
where $\chi_0^{[\mathbf{R}]}$ is the density-density response function of the inhomogeneous, free-electron gas in the static ionic configuration ${\bf R}$, and $\delta v_\mathrm{eff}$ is the effective potential
\be 
\lefteqn{\delta v_\mathrm{eff}(\mathbf{r},\omega)=\delta v_\mathrm{sc}(\mathbf{r},\omega)}&&\\
&&-\int_V d{{\bf r}_1}\int_V d{{\bf r}_2}v_\mathrm{C}(\mathbf{r}-\mathbf{r}_1)G_{ee}^{[\mathbf{R}]}(\mathbf{r}_1,\mathbf{r}_2,\omega)\delta n_e(\mathbf{r}_2,\omega)\nn
\ee 
In contrast with the previous section, here, it is the free electron response function that does not include the correlations existing  between a given electron and the other electrons of the gas.
These correlations are included through the last term of the effective potential by means of the so-called local field correction $G_{ee}^{[\mathbf{R}]}(\mathbf{r}_1,\mathbf{r}_2,\omega)$ (we extend standard definitions for the homogeneous electron gas \cite{GiulianiVignale2005,Kugler1975} to the non-homogeous case).

The response functions $\chi_\mathrm{ee}^{[\mathbf{R}]}$ and $\chi_0^{[\mathbf{R}]}$ are related through the integral equation
\be
\lefteqn{\chi_\mathrm{ee}^{[\mathbf{R}]}(\mathbf{r}_1,\mathbf{r}_2,\omega) = \chi_0^{[\mathbf{R}]}(\mathbf{r}_1,\mathbf{r}_2,\omega)}&&\label{chieechi0}\\
&&+\int_V d{{\bf r}}\int_V d{{\bf r}'}\chi_0^{[\mathbf{R}]}(\mathbf{r}_1,\mathbf{r},\omega) K^{[\mathbf{R}]} ({\bf r},{\bf r}',\omega)\chi_\mathrm{ee}^{[\mathbf{R}]}(\mathbf{r'},\mathbf{r}_2,\omega),\nn 
\ee
with the interaction kernel
\begin{equation}
  K^{[\mathbf{R}]}(\mathbf{r},\mathbf{r'},\omega) = v_{\rm C}(\mathbf{r}-\mathbf{r'})-\int_V d{{\bf r}_1} v_\mathrm{C}(\mathbf{r}-\mathbf{r}_1)G_{ee}^{[\mathbf{R}]}(\mathbf{r}_1,\mathbf{r'},\omega)\,.
\end{equation}
As shown in appendix~\ref{A:2}, with this relation, the friction coefficient (\ref{Eq:5}) can be written as the sum of two terms,
\begin{equation} \label{gammabargammadeltabargamma}
  \gamma_{\alpha\beta}^{[\mathbf{R}]} = \bar{\gamma}_{\alpha\beta}^{[\mathbf{R}]} + \delta\bar{\gamma}_{\alpha\beta}^{[\mathbf{R}]}\,.
\end{equation}
The first term writes as
\begin{equation}
  \begin{split}
    &\bar{\gamma}_{\alpha\beta}^{[\mathbf{R}]} =-\frac{1}{M}\\
    &\times\int_V d{{\bf r}}\int_V d{{\bf r}'}\bar{f}_{\alpha}^{\,\mathrm{L}}(\mathbf{r})\partial_\omega\mathrm{Im}\chi_0^{[\mathbf{R}]}(\mathbf{r},\mathbf{r'},\omega=0)\bar{f}_{\beta}^{\,\mathrm{R}}(\mathbf{r'}),
  \end{split}
\end{equation}
the spatial derivative of the effective screened potentials are given respectively by
\begin{eqnarray}
  \bar{f}_\alpha^{\,\mathrm{L}}(\mathbf{r}_1) & = & \int_V d{{\bf r}} f_\alpha(\mathbf{r})\bar{\varepsilon}_\mathrm{L}^{\,[\mathbf{R}]}(\mathbf{r},\mathbf{r}_1,\omega=0)^{-1},\\
  \bar{f}_\alpha^{\,\mathrm{R}}(\mathbf{r}_1) & = & \int_V d{{\bf r}}\bar{\varepsilon}_\mathrm{R}^{\,[\mathbf{R}]}(\mathbf{r}_1,\mathbf{r},\omega=0)^{-1} f_\alpha(\mathbf{r}),
\end{eqnarray}
where the electronic screening effect is described by the inverse of the (electron-test charge) dielectric functions
\be
  \bar{\varepsilon}_\mathrm{L}^{\,[\mathbf{R}]}(\mathbf{r},\mathbf{r}_1,\omega) = \delta(\mathbf{r}-\mathbf{r}_1)-\int_V d{{\bf r}'} \chi_0^{[\mathbf{R}]}(\mathbf{r},\mathbf{r'},\omega)K^{[\mathbf{R}]}(\mathbf{r'},\mathbf{r}_1,\omega)\nn\\ 
  \bar{\varepsilon}_\mathrm{R}^{\,[\mathbf{R}]}(\mathbf{r},\mathbf{r}_1,\omega) = \delta(\mathbf{r}-\mathbf{r}_1)-\int_V d{{\bf r}'} K^{[\mathbf{R}]}(\mathbf{r},\mathbf{r'},\omega)\chi_0^{[\mathbf{R}]}(\mathbf{r'},\mathbf{r}_1,\omega)\,.\nn
\ee
By comparison with Eqs.(\ref{g_Kubo}), we see that $\bar{\gamma}_{\alpha\beta}^{[\mathbf{R}]}$ can also be expressed as follows
\ben
\bar{\gamma}_{\alpha\beta}^{[\mathbf{R}]}&=&\frac{1}{2Mk_BT_e}\mathrm{Re}\int_0^\infty dt\left\langle\,\delta{\bar{f}}_\alpha^L(t)\delta{\bar{f}}_\alpha^R(0)\,\right\rangle_\mathrm{e}
\een
in terms of the corrrelation function $\left\langle\,\delta{\bar{f}}_\alpha^L(t)\delta{\bar{f}}_\alpha^R(0)\,\right\rangle_\mathrm{e}$ of the electron-ion force screened by the  electron-test charge dielectric functions $\bar{\varepsilon}_\mathrm{L,R}$, where ${\bar{f}}_\alpha^{L,R}(t)=e^{i\hat{h}_\mathrm{e}^0 t/\hbar} \bar{f}_\alpha^{\,\mathrm{L,R}} e^{-i\hat{h}_\mathrm{e}^0 t/\hbar}$ and $\hat{h}_e^0=\frac{\hat{{\bf p}}^2}{2m_e}+\sum_{I=1}^{N}{v_{ie}(\hat{\bf r}-{\bf R}_I)}$ is the Hamiltonian of a free electron moving in the background of the ionic potential.

The second term in Eq.(\ref{gammabargammadeltabargamma}) writes as
\begin{equation}
  \begin{split}
    &\delta\bar{\gamma}_{\alpha\beta}^{[\mathbf{R}]} =-\frac{1}{M}\\
    &\times\mathrm{Im}\int_V d{{\bf r}}\int_V d{{\bf r}'}n_\alpha'(\mathbf{r})\partial_\omega K^{[\mathbf{R}]}(\mathbf{r},\mathbf{r'},\omega=0)n_\beta'(\mathbf{r'}),
  \end{split}
\end{equation}
where $n_\alpha'(\mathbf{r}) = \int_V d{{\bf r}'}f_\alpha(\mathbf{r'})\chi_\mathrm{ee}^{[\mathbf{R}]}(\mathbf{r'},\mathbf{r},\omega=0)$.

We expect that for many systems the second term will be a small correction to the first term.
In practice, the frequency dependence of the local field correction is poorly known and the so-called static approximation $G_{ee}^{[\mathbf{R}]}(\mathbf{r}_1,\mathbf{r}_2,\omega)=G_{ee}^{[\mathbf{R}]}(\mathbf{r}_1,\mathbf{r}_2,\omega=0)$ is commonly used; in this so-called static local-field correction approximation, the last term cancels out and $\gamma_{\alpha\beta}^{[\mathbf{R}]}=\bar{\gamma}_{\alpha\beta}^{[\mathbf{R}]}$.

With Eq.(\ref{gammabargammadeltabargamma}), the electron-ion coupling factor becomes
\begin{widetext}
\begin{eqnarray}
g(T_e,T_i)&=&-\frac{3 k_\mathrm{B}n_\mathrm{i}}{M}\left\langle\frac{1}{3N}\sum_{\alpha=1}^{3N}
\mathrm{Im}\iint_V\!d{{\bf r}}d{{\bf r}'}\bar{f}_{\alpha}^{\,\mathrm{L}}(\mathbf{r})\partial_\omega\chi_0^{[\mathbf{R}]}(\mathbf{r},\mathbf{r'},\omega=0)\bar{f}_{\beta}^{\,\mathrm{R}}(\mathbf{r'}) \right\rangle\nn\\
&&\hspace{1cm}-\frac{3 k_\mathrm{B}n_\mathrm{i}}{M}\left\langle\frac{1}{3N}\sum_{\alpha=1}^{3N}
\mathrm{Im}\iint_V\!d{{\bf r}}d{{\bf r}'}n_\alpha'(\mathbf{r})\partial_\omega K^{[\mathbf{R}]}(\mathbf{r},\mathbf{r'},\omega=0)n_\alpha'(\mathbf{r'})\right\rangle\,. \label{g_noninteracting}
\end{eqnarray}
\end{widetext}

\subsection{Relation to the Kohn-Sham response function}

The previous expressions in terms of the proper response function or in terms of the local field corrections are useful for theoretical analysis using many-body techniques and for the developments of practical models (see, e.g., Sec.~\ref{section_4}).
Another approach consists in expressing Eq.(\ref{g_Kubo}) in terms of quantities directly accessible to ab-initio simulations.
The formulation in terms of the Kohn-Sham response function described here is at the basis of the first-principles molecular dynamics simulations presented in \cite{SimoniDaligaultPRL2019}.
We restrict ourselves to the presentation of the exact expressions; the practical algortihms used to numerically evaluate them will be discussed at length elsewhere \cite{SimoniDaligaultQMDpaper}.

The most important among these methods is Density Functional Theory (DFT) \cite{Ulrich2012}, in which the density deviation Eq.(\ref{delta_ne}) can be written as
\ben
\delta n_e({\bf r},\omega)&=&\int{d{\bf r}^\prime \chi_{\rm KS}^{[\mathbf{R}]} ({\bf r},{\bf r}';\omega)\delta v_{\rm eff}^{\rm KS}({\bf r}',\omega)}
\een
where the effective interaction
\be 
\delta v_\mathrm{eff}^{\rm KS}(\mathbf{r},\omega)&=&\delta v_\mathrm{ext}(\mathbf{r},\omega) + \int_V d{{\bf r}_1}v_\mathrm{C}(\mathbf{r}-\mathbf{r}_1)\delta n_e(\mathbf{r}_1,\omega)\nn\\
&&\hspace{0.75cm}+\int_V d{{\bf r}_1}f_{xc}^{[\mathbf{R}]}(\mathbf{r},\mathbf{r}_1,\omega)\delta n_e(\mathbf{r}_1,\omega)\nn\\
&=&\delta v_\mathrm{sc}(\mathbf{r},\omega) +\int_V d{{\bf r}_1}f_{xc}^{[\mathbf{R}]}(\mathbf{r},\mathbf{r}_1,\omega)\delta n_e(\mathbf{r}_1,\omega)\nn
\ee 
is the sum of the external perturbation $\delta v_{\rm ext}$ and of the linearized Kohn-Sham potential.
The latter consists of the sum of the linearized Hartree potential and of the linearized exchange-correlation potential, where  $f_\mathrm{xc}^{[\mathbf{R}]}[n_e](\mathbf{r},\mathbf{r'},\omega)$ is the so-called exchange-correlation kernel \cite{Ulrich2012}.
The Kohn-Sham density-density response function is given by \cite{Ulrich2012}
\begin{equation}
  \begin{split}
&\chi_{\rm KS}^{[\mathbf{R}]} ({\bf r}_1,{\bf r}_2;\omega)\\
&=\lim_{\eta\rightarrow 0^+}\sum_{i\neq j}{\frac{p_i^{\rm
            eq}-p_j^{\rm
            eq}}{\hbar\omega+\epsilon_i-\epsilon_j+i\eta}\mel{i}{\hat{n}_{\rm
            e}({\bf r}_1)}{j} \mel{j}{\hat{n}_{\rm e}({\bf r}_2)}{i}}
  \end{split}
\end{equation}
where $p_i^{\rm eq}=1/\left[1+e^{-(\mu(T_e)-\epsilon_i)/k_BT_e}\right]$ is the Fermi-Dirac function at temperature $T_e$,and $\ip{\bf r}{i}=\phi_i(\mathbf{r})$ are the Kohn-Sham wave functions, obtained from the solution of the following set of equations
\begin{equation}\label{Eq:6}
\Big[-\frac{\hbar^2}{2m}\nabla^2 + v_\mathrm{KS}(\mathbf{r})\Big]\phi_i(\mathbf{r}) = \epsilon_i\phi_i(\mathbf{r})\,
\end{equation}
where
\begin{equation}\label{v_KS}
v_\mathrm{KS}[n_{\rm e}]({\bf r})= \sum_{I=1}^{N}{v_{ie}({{\bf r}}-{\bf R}_I)}+
  v_\mathrm{H}[n_{\rm e}]({\bf r}) + v_\mathrm{xc}[n_{\rm e}]({\bf r}) 
\end{equation}
is the Kohn-Sham potential, $v_\mathrm{H}[n_{\rm e}](\mathbf{r})=e^2\int_V d{{\bf r}'}\frac{n_\mathrm{e}(\mathbf{r'})}{|\mathbf{r}-\mathbf{r'}|}$ is the Hartree potential, $v_\mathrm{xc}[n_{\rm e}](\mathbf{r})$ is the exchange-correlation potential, and 
\begin{equation}\label{Eq:7}
 n_\mathrm{e}(\mathbf{r}) = \sum_i p_i^{\rm eq}|\phi_i(\mathbf{r})|^2\,,
\end{equation}
is the ground state electron density of the physical system.
The present approach differs from the one of the
  previous section in the fact that, unlike $\chi_0^{[\mathbf{R}]}$
  that describes the response of a free electron gas to the potential
  $\delta v_{\rm{eff}}^{[{\bf R}]}({\bf r})$, the Kohn-Sham response function is
  the response of a system of free particles moving under the effect
  of the self-consistent Kohn-Sham potential $\delta
  v_\mathrm{KS}({\bf r})$, Eq.(\ref{v_KS}).

The response functions $\chi_\mathrm{ee}^{[\mathbf{R}]}$ and $\chi_\mathrm{KS}^{[\mathbf{R}]}$ are related through the integral equation \cite{Ulrich2012}
\be
\lefteqn{\chi_\mathrm{ee}^{[\mathbf{R}]}(\mathbf{r}_1,\mathbf{r}_2,\omega) = \chi_\mathrm{KS}^{[\mathbf{R}]}(\mathbf{r}_1,\mathbf{r}_2,\omega)}&&\label{DysonchichiKS}\\
&&+\int_V d{{\bf r}}\int_V d{{\bf r}'}\chi_\mathrm{KS}^{[\mathbf{R}]}(\mathbf{r}_1,\mathbf{r},\omega)K_\mathrm{KS}^{[\mathbf{R}]}(\mathbf{r},\mathbf{r'},\omega)\chi_\mathrm{ee}^{[\mathbf{R}]}(\mathbf{r'},\mathbf{r}_2,\omega),\nn
\ee
where the kernel $K_\mathrm{KS}^{[\mathbf{R}]}(\mathbf{r},\mathbf{r'},\omega)$ is now given by the sum of the Hartree and exchange-correlation contributions
\be 
  K_\mathrm{KS}^{[\mathbf{R}]}(\mathbf{r},\mathbf{r'},\omega) = v_{\rm C}(\mathbf{r}-\mathbf{r'}) + f_\mathrm{xc}^{[\mathbf{R}]}[n_\mathrm{e}](\mathbf{r},\mathbf{r'},\omega),
\ee 

Using Eq.~(\ref{DysonchichiKS}) into Eq.~(\ref{Eq:5}),  each friction coefficient becomes the sum of two contributions,
\begin{equation}\label{Eq:8a}
  \gamma_{\alpha\beta}^{[\mathbf{R}]}=\tilde{\gamma}_{\alpha\beta}^{[\mathbf{R}]}+\delta{\tilde{\gamma}}_{\alpha\beta}^{[\mathbf{R}]}
\end{equation}
The first term is
\be
\lefteqn{\tilde{\gamma}_{\alpha\beta}^{[\mathbf{R}]} =-\frac{1}{M}}&& \label{Eq:8}\\
&\times&\mathrm{Im}\int_V d{{\bf r}_1}\int_V d{{\bf r}_2}\tilde{f}_\alpha^{\,\mathrm{L}}(\mathbf{r}_1)\partial_\omega\chi_\mathrm{KS}^{[\mathbf{R}]}(\mathbf{r}_1,\mathbf{r}_2,\omega=0)\tilde{f}_\beta^{\,\mathrm{R}}(\mathbf{r}_2), \nn
\ee
where the spatial derivative of the effective screened potentials are given respectively by
\ben 
  \tilde{f}_\alpha^{\,\mathrm{L}}(\mathbf{r}_1) & = & \int_V d{{\bf r}} f_\alpha(\mathbf{r})\tilde{\varepsilon}_\mathrm{L}^{\,[\mathbf{R}]}(\mathbf{r},\mathbf{r}_1,\omega=0)^{-1},\\
  \tilde{f}_\alpha^{\,\mathrm{R}}(\mathbf{r}_1) & = & \int_V d{{\bf r}}\tilde{\varepsilon}_\mathrm{R}^{\,[\mathbf{R}]}(\mathbf{r}_1,\mathbf{r},\omega=0)^{-1} f_\alpha(\mathbf{r}),
\een 
and the screening effect of the electronic subsystem is described by the inverse of the dielectric functions
\ben
  \tilde{\varepsilon}_\mathrm{L}^{\,[\mathbf{R}]}(\mathbf{r},\mathbf{r}_1,\omega)= \delta(\mathbf{r}-\mathbf{r}_1)-\int_V d{{\bf r}'} \chi_\mathrm{KS}^{[\mathbf{R}]}(\mathbf{r},\mathbf{r'},\omega)K_\mathrm{KS}^{[\mathbf{R}]}(\mathbf{r'},\mathbf{r}_1,\omega)\\ 
  \tilde{\varepsilon}_\mathrm{R}^{\,[\mathbf{R}]}(\mathbf{r},\mathbf{r}_1,\omega)= \delta(\mathbf{r}-\mathbf{r}_1)- \int_V d{{\bf r}'} K_\mathrm{KS}^{[\mathbf{R}]}(\mathbf{r},\mathbf{r'},\omega)\chi_\mathrm{KS}^{[\mathbf{R}]}(\mathbf{r'},\mathbf{r}_1,\omega)
\een
By comparison with Eqs.(\ref{g_Kubo}), we see that $\tilde{\gamma}_{\alpha\beta}^{[\mathbf{R}]}$ can also be expressed as follows
\be
\tilde{\gamma}_{\alpha\beta}^{[\mathbf{R}]}&=&\frac{1}{2Mk_BT_e}\mathrm{Re}\int_0^\infty dt\left\langle\,\delta{\tilde{f}}_\alpha^L(t)\delta{\tilde{f}}_\alpha^R(0)\,\right\rangle_\mathrm{e}\,,\label{tildegammaab}
\ee
in terms of the corrrelation function $\left\langle\,\delta{\tilde{f}}_\alpha^L(t)\delta{\tilde{f}}_\alpha^R(0)\,\right\rangle_\mathrm{e}$ of the KS particle-ion force screened by the KS particle -test charge dielectric function $\tilde{\varepsilon}_\mathrm{L,R}$, where ${\tilde{f}}_\alpha^{L,R}(t)=e^{i\hat{h}_\mathrm{KS} t/\hbar} \tilde{f}_\alpha^{\,\mathrm{L,R}} e^{-i\hat{h}_\mathrm{KS} t/\hbar}$, and $\hat{h}_{KS}=\frac{\hat{{\bf p}}^2}{2m_e}+v_{KS}(\hat{\bf r})$ is the KS Hamiltonian.

The second term in the expression (\ref{Eq:8a}), instead, represents a quantum many body correction including all the memory effects
\be
\lefteqn{\delta{\tilde{\gamma}}_{\alpha\beta}^{[\mathbf{R}]} =-\frac{1}{M}}&& \label{Eq:8b}\\
&\times& \mathrm{Im}\int_V d{{\bf r}_1}\int_V d{{\bf r}_2}n_\alpha'(\mathbf{r}_1)\partial_\omega f_\mathrm{xc}^{[\mathbf{R}]}(\mathbf{r}_1,\mathbf{r}_2,\omega=0)n_\beta'(\mathbf{r}_2),\nn
\ee
where $n_\alpha'(\mathbf{r})= \int_V d{{\bf r}'}f_\alpha(\mathbf{r'})\chi_\mathrm{ee}^{[\mathbf{R}]}(\mathbf{r'},\mathbf{r},\omega=0)$.
In practice \cite{SimoniDaligaultQMDpaper}, the frequency dependence of the exchange-correlation kernel remains poorly known \cite{Nazarov2009,Thiele2014}; we expect it to be a small correction to the first term for many systems.

We note that the sum rules (\ref{Eq:9})-(\ref{Eq:9p}) discussed in Sec.~\ref{Kubo_relations} become (see Appendix~\ref{A:2})
\be
 \sum_\mathrm{I,J=1}^{N}{\tilde{\gamma}_{Ix,Jy}^{[\mathbf{R}]}}=0\quad,\quad
 \sum_\mathrm{I,J=1}^{N}{\delta{\tilde{\gamma}}_{Ix,Jy}^{[\mathbf{R}]}}=0 
\ee
for all x,y.
In practice, these sum rules provide a useful test of the accuracy or the consistency of the numerical calculations.

By using Eq.(\ref{Eq:8a}), the electron-ion coupling factor becomes
\begin{widetext}
\begin{eqnarray}
g(T_e,T_i)&=&-\frac{3 k_\mathrm{B}n_\mathrm{i}}{M}\left\langle\frac{1}{3N}\sum_{\alpha=1}^{3N}
\mathrm{Im}\iint_V\!d{{\bf r}_1}d{{\bf r}_2}\tilde{f}_\alpha^{\,\mathrm{L}}(\mathbf{r}_1)\partial_\omega\chi_\mathrm{KS}^{[\mathbf{R}]}(\mathbf{r}_1,\mathbf{r}_2,\omega=0)\tilde{f}_\beta^{\,\mathrm{R}}(\mathbf{r}_2)
\right\rangle\nn\\
&&\hspace{1cm}-\frac{3 k_\mathrm{B}n_\mathrm{i}}{M}\left\langle\frac{1}{3N}\sum_{\alpha=1}^{3N}
\mathrm{Im}\iint_V\!d{{\bf r}_1}d{{\bf r}_2}n_\alpha'(\mathbf{r}_1)\partial_\omega f_\mathrm{xc}^{[\mathbf{R}]}(\mathbf{r}_1,\mathbf{r}_2,\omega=0)n_\beta'(\mathbf{r}_2) \label{g_kohnsham}
\right\rangle
\end{eqnarray}
\end{widetext}

We find important to remark that Eq.(\ref{Eq:8a}) is an exact rewriting of the Kubo formula (\ref{gammaab_correlation}).
It does not correspond to the Kubo-Greenwood approximation that is widely used to calculate other transport properties such as the electrical and thermal conductivities.
In the Kubo-Greenwood (KG) approximation, the Kohn-Sham wave functions are assumed to represent the single-particle excitations of the electronic system, i.e. the many-body electronic states are approximated by Slater determinant of KS orbitals.
Here, this approximation amounts to set $\chi_\mathrm{ee}^{[\mathbf{R}]}$ in Eq.(\ref{Eq:5}) equal to $\chi_\mathrm{KS}^{[\mathbf{R}]}$, yielding
\be
\gamma_{\alpha\beta}^{[\mathbf{R}]}&\simeq& -\frac{1}{M}\!\iint_V\!d{{\bf r}_1}d{{\bf r}_2}f_\alpha(\mathbf{r}_1)\partial_\omega\mathrm{Im}\chi_\mathrm{KS}^{[\mathbf{R}]} (\mathbf{r}_1,\mathbf{r}_2,0)f_\beta(\mathbf{r}_2)\nn\\
&=&\frac{1}{2Mk_BT_e}\mathrm{Re}\int_0^\infty dt\left\langle\,\delta\hat{f}_\alpha^{KG}(t)\delta\hat{f}_\alpha^{KG}(0)\,\right\rangle_\mathrm{e} \label{gammaabGKapproximation}
\ee
which should be compared with Eq.(\ref{Eq:8a}).
Here $\hat{f}_\alpha^{KG}(t)=-e^{i\hat{h}_\mathrm{KS} t/\hbar} \partial_{R_\alpha}\hat{V}_{ie}e^{-i\hat{h}_\mathrm{KS} t/\hbar}$ is the bare electron-ion force propagated by the Kohn-Sham Hamiltonian.
Unlike the force $\tilde{f}_\alpha$ in Eq.(\ref{tildegammaab}), $\hat{f}_\alpha^{KG}$ fails to include the effects of the electronic screening on the electron-ion interactions.
As a consequence, the KG approximation (\ref{gammaabGKapproximation}) can be shown to diverge logarithmically at large interparticle distances $|{\bf r}_1-{\bf r}_2|$.
This is fully analogous to the well-known infrared divergence that occurs in the basic calculations of scattering cross sections in plasma physics when screening effects are neglected \cite{Spitzer1962}.

\section{Relation to other models} \label{section_4}

The electron-ion coupling formula (\ref{g_Kubo}) includes self-consistently the quantum mechanical and statistical nature of electrons, the thermal effects and the correlations between all particles.
In this section we show that our theory reduces to well-known models developed in plasma and condensed matter physics when one or several of these effects are treated approximately or neglected.

We begin in Sec.~\ref{section_4_1} and \ref{section_4_2} with discussing the reduction to the Spitzer formula in the hot plasma limit \cite{Spitzer1962}, to the Fermi golden rule formula in the limit of weak electron-ion interactions \cite{Hazak2001}, and to the model developed by Daligault and Dimonte that includes important electron-ion correlation effects \cite{DaligaultDimonte2009}.
We then explain that the problem of the calculation of the electron-ion coupling is particularly well adapted to average atom model calculations.
In Sec.~\ref{section_4_3}, we show that our theory also applies to hot solids, namely to solid metals with lattice and electronic temperatures much greater than the Debye temperature.
When the ionic motions are described within the phonon approximation, our theory reproduces the standard electron-phonon coupling formula \cite{Allen1987} in the high temperature limit; the full formula includes effects beyond the harmonic approximation.
Finally, in Sec.~\ref{section_4_4}, we relate our theory to a simplified model of the electon-phonon coupling factor due to Wang et al. \cite{Wang1994}.

In the following, $v_{ie}(k)=\int{\! d{\bf r} v_{ie}(r)e^{-i{\bf k}\cdot{\bf r}}}$ and $v_C(k)=4\pi e^2/k^2$ are the spatial Fourier transforms of the electron-ion and Coulomb interaction potentials.

\subsection{Reduction to the Spitzer model and the Fermi golden rule formula} \label{section_4_1}

In the general formula (\ref{g_Kubo}), the effects of the interactions between the ionic and electronic subsystems are included non-perturbatively.
Simpler models are obtained if one assumes that the interaction $v_{ie}$ is weak.
Then, to lowest order in $v_{ie}$,  the response function $\chi_{ee}^{[{\bf R}]}$ can be approximated by the density-density response function $\chi_{jel}$ of the homogeneous electron gas (a.k.a. jellium) at temperature $T_e$, i.e.
\ben
\chi_{ee}^{[{\bf R}]}({\bf r},{\bf r}',\omega)\approx\chi_{jel}(|{\bf r}-{\bf r}'|,\omega)\,.
\een
With this approximation, the electron-ion coupling formula (\ref{g_Kubo_b}) becomes
\be
\lefteqn{g(T_e,T_i)=-\frac{k_\mathrm{B}n_\mathrm{i}}{M}}&& \label{g_weak_vie_1}\\
&&\times\iint_V\!d{{\bf r}_1}d{{\bf r}_2} \vec{\nabla} v_{ie}({\bf r}_1)\cdot \vec{\nabla} v_{ie}({\bf r}_2)\, \partial_\omega \mathrm{Im}\chi_{jel}(|\mathbf{r}_1-\mathbf{r}_2|,0)\nn
\ee
In the thermodynamic limit, this expression is conveniently rewritten such as
\be
g(T_e,T_i)=-\frac{k_\mathrm{B}n_\mathrm{i}}{2\pi^2M}\int_{0}^{\infty}{dk\left|v_{ie}(k)\right|^2 k^4 \partial_\omega \mathrm{Im}\chi_{jel}(k,0)}\,,\nn\\\label{gchieeapprox}
\ee
where $\chi_{jel}(k,\omega)=\int{\! d{\bf r}\chi_{jel}(r,\omega)e^{-i{\bf k}\cdot{\bf r}}}$.
Similarly, the formulas (\ref{g_proper}) and (\ref{g_noninteracting}) become
\be
\lefteqn{g(T_e,T_i)}&&\label{gchi0approx1}\\
&=&-\frac{k_\mathrm{B}n_\mathrm{i}}{2\pi^2M}\int_{0}^{\infty}{dk\left|\frac{v_{ie}(k)}{1-v_C(k)\tilde{\chi}(k,0)}\right|^2k^4 \partial_\omega \mathrm{Im}\tilde{\chi}(k,0)}\nn
\ee
and
\be
\lefteqn{g(T_e,T_i)}&&\label{gchi0approx2}\\
&=&-\frac{k_\mathrm{B}n_\mathrm{i}}{2\pi^2M}\int_{0}^{\infty}{dk\left|\frac{v_{ie}(k)}{1-v_C(k)\left[1-G_{ee}(k,0)\right]\chi_{0}(k,0)}\right|^2}\nn\\
&&\quad\times\, k^4\left[\partial_\omega \mathrm{Im}\chi_0(k,0)+|\chi_0(k,0)|^2 v_C(k)\partial_\omega G_{ee}(k,0)\right]\,.\nn
\ee
where $\tilde{\chi}$, $\chi_0$ and $G_{ee}$ are the proper response function, the non-interacting (a.k.a. Lindhard) response function and the local field correction of the jellium model \cite{GiulianiVignale2005}.
Equations (\ref{gchieeapprox}), (\ref{gchi0approx1}) and (\ref{gchi0approx2}) correspond to the so-called `Fermi golden rule formula' for the electron-ion coupling derived in Ref.~\cite{Hazak2001} by first calculating the energy exchanges between the electron and ion subsytems within the framework of linear response theory and then by taking the small ion to electron velocity ratio into account.

The relation of Eq.(\ref{gchieeapprox}) to the celebrated Spitzer formula was discussed elsewhere (e.g., see \cite{DaligaultDimonte2009}), and we only briefly recall the result for completeness.
When electron-electron correlation effects are neglected, $\tilde\chi=\chi_0$ and $G_{ee}= 0$, and
\be
\lefteqn{g(T_e,T_i)=-\frac{k_\mathrm{B}n_\mathrm{i}}{2\pi^2M}}&&\label{g_towards_plasmas}\\
&&\times\int_{0}^{\infty}{\!dk\left|\frac{v_{ie}(k)}{1-v_C(k)\chi_{0}(k,0)}\right|^2 \!k^4\partial_\omega \mathrm{Im}\chi_0(k,0)}\nn
\ee
The familar plasma physics results are recovered using the Coulomb interaction $v_{ie}(r)=-Ze^2/r$ in Eq.(\ref{g_towards_plasmas}), yielding
\be
g(T_e,T_i)=4n_i Z^2\frac{(2\pi m_eM)^{1/2}}{(Mk_BT_e)^{3/2}}\ln\Lambda \label{g_almost_Spitzer}
\ee
in terms of the Coulomb logarithm
\be
\ln\Lambda=\int_0^\infty{\frac{dk}{k}\frac{k^4}{\left(k^2+\lambda_{sc}^{-2}\right)^2}f(k/2)} \label{Coulomblog}
\ee
with the screening length $\lambda_{sc}=1/\sqrt{4\pi e^2\chi_0(k,0)}$ and $f(k)$ is the Fermi-Dirac function (see Eq.(\ref{fofk})).
In the classical limit $\hbar\to 0$, $f(k)=1$ and $\lambda_{sc}=\sqrt{k_BT_e/4\pi n_e e^2}$ is the Debye-H{\"u}ckel screening length, and Eq.(\ref{g_almost_Spitzer}) becomes the celebrated Spitzer formula, which logarithmically diverges at large $k$.
In the non-degenerate limit, Eq.(\ref{Coulomblog}) becomes $\ln\Lambda=\int_0^\infty{\frac{dk}{k}\frac{k^4}{(k^2+\lambda_{De}^{-2})^2}e^{-\lambda_e^2 k^2/8}}$ is convergent, where  $\lambda_e=\hbar/\sqrt{m_ek_BT_e}$ is the electronic thermal de Broglie wavelength.

\subsection{Going beyond the weak electron-ion interaction approximation using average atom models} \label{section_4_2}

In the previous section, we have discussed the general expression (\ref{g_Kubo}) in the limit of weak electron-ion interactions.
Here, we discuss approaches to go beyond this approximation.
To this end, we remark that the formula (\ref{Eq:2f}) can be written as
\be
g(T_\mathrm{e},T_\mathrm{i})&=&3 k_\mathrm{B}n_\mathrm{i} \Gamma(T_e,T_i)\,,\label{g_gamma}
\ee
where
\be
\Gamma(T_e,T_i)=\bigg\langle\frac{1}{3N}\sum_{\alpha=1}^{3N}\gamma_{\alpha\alpha}^{[\mathbf{R}]}(T_e,T_i)\bigg\rangle \label{average_friction}
\ee
can be regarded as the averaged friction coefficient felt by any ion in the system along any direction of motion.

A natural approximation for $\Gamma$ consists in identifying it with the friction felt by a single impurity embedded in an electron gas (jellium) or by a slow projectile (i.e., whose velocity is much smaller than the electronic velocities),
\be
\Gamma=\frac{1}{6Mk_BT_e}\mathrm{Re}\int_0^\infty dt\langle\,\delta\hat{{\bf f}}(t)\cdot\delta\hat{{\bf f}}(0)\,\rangle_\mathrm{e} \label{Gamma_Kubo}
\ee
where ${\bf f}$ is the force between the impurity and the electrons.
This problem has been extensively studied in the past (see, e.g. \cite{dAgliano1975}) and, for completeness, we recall results useful to the present work.
When the electron-impurity interaction is treated within the framework of linear response,
\be
\Gamma=-\frac{1}{6\pi^2 M}\int_{0}^{\infty}{dk\left|v_{ie}(k)\right|^2k^4 \partial_\omega \mathrm{Im}\chi_{jel}(k,0)}\,. \label{Gamma_linear}
\ee
When used in Eq.(\ref{g_gamma}), we, not surprisingly, retrieve the Fermi golden rule formula (\ref{gchieeapprox}) discussed above.

As suggested in Ref.~\cite{Dufty1996}, the so-called `disconnected approximation' can be used to extend Eq.(\ref{Gamma_linear}) beyond the weak electron-ion interaction approximation.
This approximation, which orginates from works in the classical kinetic theory of strongly coupled plasmas \cite{Gould1977,Boercker1982}, neglects the effect of the slow impurity on the electron dynamics, but accounts for the average distortion of the electronic density around the impurity.
It amounts to replacing in Eq.(\ref{Gamma_linear}) the term $v_{ie}(k)^2$ by $v_{ie}(k)^2 [1-G_{ie}(k)]$, where $G_{ie}$ is the electron-ion local field correction $G_{ie}(k)$, yielding
\be
\lefteqn{g(T_e,T_i)}&&\label{g_DaligaultDimonte}\\
&&=-\frac{k_\mathrm{B}n_\mathrm{i}}{2\pi^2 M}\int_{0}^{\infty}{\!\!dk\left|\frac{v_{ie}(k)}{1\!-\!v_C(k)\!\left[1\!-\!G_{ee}(k,0)\right]\!\chi_{0}(k,0)}\right|^2}\nn\\
&&\hspace{2.5cm}\times \left[1-G_{ie}(k)\right]k^4\partial_\omega \mathrm{Im}\chi_0(k,\!0)\nn
\ee
This corresponds to the model derived by Daligault and Dimonte in \cite{DaligaultDimonte2009} using a very different method.
We refer to Ref.~\cite{DaligaultDimonte2009} for a detailed discussion of Eq.(\ref{g_DaligaultDimonte}).

Let assume for the moment that electron interactions can be neglected ($V_{ee}\equiv 0$).
Electrons then move independently in the potential of the ionic impurity and the Kubo relation (\ref{Gamma_Kubo}) can then be expressed in terms of the basic scattering properties of the electron-ion potential $v_{ie}$ (see e.g. \cite{dAgliano1975}), which gives
\be
\lefteqn{g(T_e,T_i)=\frac{\hbar^2 n_\mathrm{i}}{\pi^2 m_e M T_e}} \label{g_binary}&&\\
&&\hspace{1.cm}\times\int_0^\infty{dk\,k^5n_{\rm FD}(\epsilon_k)\left(1-n_{\rm FD}(\epsilon_k)\right) \sigma_{tr}(k)}\nn
\ee
where $\epsilon_k=\hbar^2k^2/2m_e$, $n_{\rm FD}(\epsilon)=1/(1+e^{-(\mu-\epsilon)/k_BT_e})$ is the Fermi-Dirac distribution, and 
$\sigma_{tr}$ is the cross section for binary collisions
\be
\sigma_{tr}(k)=\frac{4\pi}{k^2}\sum_l{(l+1)\sin^2\left(\delta_l(k)-\delta_{l+1}(k)\right)}\,, \label{cross_section}
\ee
where $\delta_l(k)$ is the phase shift of the $l^{\rm th}$ partial wave at momentum $\hbar k$ calculated for the spherically symmetric $v_{ie}(r)$ (see appendix \ref{appendix_phase_shifts}).
This formula is applicable to any temperature $T_e$.
In particular, at $T_e=0$,
\be
g(T_e,T_i)=\frac{\hbar k_\mathrm{B}n_\mathrm{i} k_F^4}{\pi^2 M} \sigma_{tr}(k_F)\,, \label{g_binary_0}
\ee
and, at high $T_e$, the result (\ref{g_binary}) agrees with that obatined from the classical Boltzmann-Lorentz kinetic theory \cite{FerzigerKaper1972}
\be
g(T_e,T_i)=\frac{ 8\pi k_\mathrm{B}n_\mathrm{i} n_e m_e}{M}(2\pi m_ek_BT_e)^3\,\Omega^{(1,1)} \label{g_binary_nondeg}
\ee
with
\be
\Omega^{(1,1)}=\sqrt{\frac{k_BT_e}{2\pi m_e}}\int_0^{\infty}{d\gamma e^{-\gamma^2} \sigma_{tr}\left(\frac{\sqrt{2m_ek_BT_e}}{\hbar}\gamma\right)}
\ee
The result (\ref{g_binary}) can be effectively used in conjunction with an average atom model to include the effects of electron and ion interactions that affect the electron-ion cross section in a plasma.
Average atom models have been a quite popular approximate method to model both the equation-of-state and transport properties of dense ionized matter \cite{Faussurier2010,Faussurier2016,StarrettSaumon2013,Starrett2017}.
They have proved to be accurate enough to be useful while being computationally much more expedient.
An average atom model assumes that the physical system is spherically symmetric about a central nucleus and one calculates with finite-temperature density functional theory the electronic structure of the central ions and of the surrounding conduction electrons.
As we have seen above the electron-ion coupling factor is related to the averaged friction coefficient felt by any ion in the system along any direction of motion; its calculation is thus particularly well adapted for this transport property. 
Many variations exist that differ in the description of the surrounding plasma, e.g. via boundary conditions or by coupling the model with the theory of fluids.
Electrons are treated as independent particles subject to the Kohn-Sham potential $v_{KS}(r)$ given by the sum of the central nuclear potential, the spherically averaged distribution of surrounding ions, and the self-consistent Hartree and exchange correlation potentials.
The expressions (\ref{g_binary}) and (\ref{cross_section}) still apply but should be computed using the phase shifts $\delta_l(k)$ corresponding to the effective potential $v_{KS}(r)$.
Results for the friction coefficient felt by a charge immersed in a homogeneous electron gas were presented in Ref.~\cite{Nazarov2005}.

\subsection{Relation to the electron-phonon coupling factor} \label{section_4_3}

In hte case of solids, the energy exchanges between electrons and ions are generally described in terms of interactions between electrons and phonons.
In the phonon approximation, the total Hamiltonian is given by $\hat{H}= \hat{H}_{\rm e}({\bf R}_0)+ \hat{H}_{\rm ph} +\hat{H}_{\rm e-ph}$, where $\hat{H}_{\rm e}({\bf R}_0)$ is the electronic Hamiltonian in the potential of the Bravais lattice (${\bf R}^0$ denote the equilibrium lattice positions), $ \hat{H}_{\rm ph}$ is the Hamiltonian of phonons, and $\hat{H}_{\rm e-ph}$ is the electron-phonon Hamiltonian (see Appendix \ref{appendix_eph}).
Here we show that, for hot solids, the standard electron-phonon coupling factor can be readily derived from the electron-ion coupling formula (\ref{g_Kubo}).
For simplicity, we use the assumption of thermal phonons characterized by a single temperature $T_i$, although recent investigations suggest that it could lead to marked disagreement with experimental observations \cite{Waldecker2016, Henighan2016,Maldonado2017,Lu1}.

Most of the works in condensed matter physics rely on a formula for the electron-ion coupling derived by Allen \cite{Allen1987}.
Here, we prefer to work with a generalization of Allen's formula that, unlike the latter, does not approximate from the outset the electrons with the Bloch states.
This generalized formula naturally reduces to Allen's formula in the appropriate limit as shown in appendix \ref{appendix_eph}.
As shown in appendix \ref{appendix_eph_1}, the rate of change of the total electron energy due to the absorption and emission of phonon excitations can be written as
\be
\frac{dE_e}{dt}\!& = &\!
4 \hbar \sum_{{\bf q}}\int_{0}^\infty\frac{d\omega}{2\pi}\sum_{\bf
      G,G'} v_{ie}({\bf q}+{\bf G})^*v_{ie}({\bf q}+{\bf G}') \nn\\
&&\hspace{0.5cm}\times \omega{\rm Im}\chi_{{\bf G},{\bf G'}}^{[{\bf R}_0]}  ({\bf q},\omega){\rm Im}\chi_{{\bf G},{\bf G'}}^{\rm
  ph}({\bf q},\omega)\label{dEdt_eph}\\
&&\hspace{0.5cm}\times \left[n_B\left(\frac{\hbar\omega}{k_BT_i}\right)-n_B\left(\frac{\hbar\omega}{k_BT_e}\right)\right]\nn
\ee
with $n_B(x)=1/(e^x-1)$.
Here $\chi_{{\bf G},{\bf G'}}^{\rm ph}({\bf q},\omega)$ is the Fourier transformed density-density response function of the ion subsystem calculated in the phonon approximation (see appendix \ref{appendix_eph_chiiiph}), and $\chi_{{\bf G},{\bf G'}}^{[{\bf R}_0]} ({\bf q},\omega)$ is the Fourier transformed density-density response function $\chi_{ee}^{[{\bf R}_0]}$ of the electron system described by the Hamiltonian $\hat{H}_e$, i.e. of the system of interacting electrons in the background of the perfect lattice potential,
\be
\lefteqn{\chi_{ee}^{[{\bf R}_0]} ({\bf r},{\bf r'},\omega)}&& \label{chi_eeph}\\
&&=\frac{1}{V}\sum_{\bf q}\sum_{\bf G,G'}e^{i({\bf q}+{\bf G})\cdot{\bf r}}e^{-i({\bf q}+{\bf G'})\cdot{\bf r'}}\chi_{{\bf GG'}}^{[{\bf R}_0]}({\bf q},\omega)\nn
\ee
As shown in appendix \ref{appendix_eph_allen}, Eq.(\ref{dEdt_eph}) reduces to Allen's work (see Eq.(6) in \cite{Allen1987}) in the limit of Bloch electrons.

Equation (\ref{dEdt_eph}) is valid for any temperatures $T_{e,i}$.
The rate of energy change (\ref{dEdt_eph}) can be simplified at high temperatures, i.e. for electronic and ion temperatures greater than the Debye temperature $\Theta_D$ ($k_B\Theta_D=\hbar\omega_D$ with the Debye frequency $\omega_D=\underset {\bfq,\lambda}{\rm max}\{\omega_{\bfq\lambda}\}$ (typically $0.01-0.04$ eV \cite{AshcroftMerminbook})).
In that limit, as shown in appendix \ref{appendix_eph_highT}, Eq.(\ref{dEdt_eph}) simplifies to
\be
\frac{1}{V}\frac{dE_e}{dt}=-g_{e-ph}\left[T_e-T_i\right] \label{dEedtphhighT}
\ee
where the electron-phonon coupling factor $g_{e-ph}$ is given by
\be
g_{e-ph}&=&
-\frac{k_B n_i}{M}\frac{1}{V}\sum_{{\bf q}}\sum_{\bf G,G'} v_{ie}({{\bf q}+{\bf G}})^*v_{ie}({{\bf q}+{\bf G'}})\nn\\
&&\hspace{0.25cm}\times\,  ({\bf q}+{\bf G})\cdot({\bf q}+{\bf G'})\,\partial_\omega {\rm Im}\chi_{{\bf GG'}}({\bf q},0) \label{geph}
\ee
Remarkably, this expression is readily obtained from our general formula (\ref{g_Kubo}) for the electron-ion coupling by substituting in Eq. (\ref{g_Kubo_b}) the expression (\ref{chi_eeph}) for the electron-electron density response function (see appendix \ref{appendix_eph_3} for the details).
Since Eq.(\ref{g_Kubo_b}) is an approximation of the density response function of electrons in a solid, the formula (\ref{g_Kubo}) extends the electron-phonon coupling formula (\ref{geph}).
In particular, it goes beyond the harmonic phonon approximation as the ionic configurations ${\bf R}$ in Eq.(\ref{g_Kubo}) include those thermally sampled that are not described by the small harmonic lattice vibrations.
Such anharmonic motions become increasingly important as one approaches melting conditions.
We refer the reader to \cite{SimoniDaligaultPRL2019} for preliminary first-principle calculations of the electron-phonon coupling based on Eq.(\ref{g_Kubo}).

\subsection{Relation to the Lin et al. model} \label{section_4_4}

We finally relate our approach to the following model due to Wang et al. \cite{Wang1994,Lin_et_al_2008},
\be
G_{\rm e-ph}\approx G_0^{\rm e-ph}\!\!\int\limits_{-\infty}^{\infty}{\left[\frac{g(\epsilon)}{g(\epsilon_F)}\right]^2\!\!\left(\!-\frac{\partial\,n_{\rm FD}(\epsilon)}{\partial \epsilon}\right) \!d\epsilon}\,, \label{Lin_et_al}
\ee
obtained as a simplification valid at high temperatures of Allen's electron-phonon coupling fomula \cite{Allen1987}.
Here $g(\epsilon)$ is the electron density of states (DOS), which is computable with DFT, and $G_0^{\rm e-ph}=\pi\hbar k_B\lambda\langle\omega^2\rangle g(\epsilon_F)$, where $\epsilon_F=k_BT_F$ is the Fermi energy, $\langle\omega^2\rangle$ is the second moment of the phonon spectrum, and $\lambda$ is the electron-phonon mass enhancement factor.
In previous works, the prefactor $G_0^{\rm e-ph} $ was either set to match an experimental measurement at low electronic temperature \cite{Lin_et_al_2008}, or was calculated ab-initio \cite{Waldecker2016, JiZhang2016}.
Although derived for crystalline solids, the model (\ref{Lin_et_al}) was used in recent works on warm dense matter systems \cite{Leguay2013,Cho_et_al_2015,Jourdain_et_al_2018,Ogitsu2018}.
Remarkably, an expression similar to Eq.(\ref{Lin_et_al}) also results from Eq.(\ref{g_kohnsham}) if one neglects the second term and if one assumes that the matrix elements between the Kohn-Sham states of the force operator $\delta \tilde{f}^{L,R}$ depend weakly on the energies and spatial directions, which yields
\be
G_{ei}\approx \left\langle G_0^{ei} \int\limits_{-\infty}^{\infty}{\left[\frac{g^{[{\bf R}]}(\epsilon)}{g^{[{\bf R}]}(\epsilon_F)}\right]^2\!\!\left(\!-\frac{\partial\,n_{\rm FD}(\epsilon)}{\partial \epsilon}\right) d\epsilon}\right\rangle\,,\label{our_Lin_et_al}
\ee
as shown in appendix~\ref{appendix_Lin_derivation}.
Here $G_0^{ei}=|\delta\tilde{f}|^2g^{[{\bf R}]}(\epsilon_F)^2$, where $g^{[{\bf R}]}(\epsilon)$ is the density of states of the Kohn-Sham system in the frozen ionic configuration ${\bf R}$, and $\delta\tilde{f}$ is the characteristic matrix element.
The formulas (\ref{Lin_et_al}) and (\ref{our_Lin_et_al}) highlight the interplay between the DOS and the distribution of electronic states, which, as shown by Lin et al. \cite{Lin_et_al_2008}, results in a strong dependence on the chemical composition and often on sharp variations with $T_e$.
In \cite{SimoniDaligaultPRL2019}, we have compared our results to predictions based on (\ref{Lin_et_al}) reported by others and on Eq.(\ref{our_Lin_et_al}) with $G_0^{ei}$ set to reproduce the value of $G_{ei}$ at the lowest $T_e$ considered.
We find that the simplified models (\ref{Lin_et_al}) and (\ref{our_Lin_et_al}) tend to overestimate the dependence on $T_e$ or predict variations at odds with the full calculation.



\acknowledgments

This work was supported by the US Department of Energy through the Los Alamos National Laboratory through the LDRD Grant No. 20170490ER and the Center of Non-Linear Studies (CNLS).
Los Alamos National Laboratory is operated by Triad National Security, LLC, for the National Nuclear Security Administration of U.S. Department of Energy (Contract No. 89233218CNA000001).


\begin{appendix}

\section{Miscellaneous properties of correlation and response functions} \label{appendix_1}

A detailed exposition of the definitions and properties recalled below can be found in standard textbooks, e.g., \cite{Kubo_book,GiulianiVignale2005}

\subsection{Quantum correlation functions}

In quantum statistical mechanics, there are several ways to measure the temporal correlations between two observables $\hat{A}$ and $\hat{B}$ at thermal equilibrium.
These include: the canonical Kubo relation
\ben 
K(t)=\frac{1}{\beta}\int_0^{\beta} d\lambda \left\langle\, e^{\lambda\hat{H}_\mathrm{e}^{[\mathbf{R}]}}\delta\hat{B} e^{-\lambda\hat{H}_\mathrm{e}^{[\mathbf{R}]}}\delta\hat{A}(t)\,\right\rangle
\een 
the symmetrized correlation function,
\ben
S(t)=\frac{1}{2}\left\langle\, \delta\hat{A}(t) \delta\hat{B}+\delta\hat{B} \delta\hat{A}(t)\,\right\rangle\,,
\een
and the unsymmetrized correlation function,
\ben
C(t)=\left\langle\,\delta\hat{A}(t) \delta\hat{B}\,\right\rangle\,,
\een
where in this appendix $\langle\dots\rangle={\rm Tr}\left(e^{-\beta\hat{H}}...\right)/{\rm Tr e^{-\beta\hat{H}}}$ indicates a canonical thermal average.
In the classical limit, the three definitions are equivalent.

{\it Lehmann representation.} 
By expanding over the eigenspectrum of the Hamiltonian $\hat{H}$, $\hat{H}|n\rangle=E_n|n\rangle$, the Fourier transforms of the time correlation functions can be written as
\be 
\lefteqn{K(\omega)=-\frac{2\pi \hbar}{\beta}}&& \label{K_lehmann}\\
&\times&\sum_{n,m}{\frac{P_n^{eq}-P_m^{eq}}{E_n-E_m} \langle n|\delta\hat{A} |m\rangle \langle m|\delta\hat{B} |n\rangle\delta(E_n-E_m-\hbar\omega)}\nn\\
\lefteqn{S(\omega)=\pi \hbar}&&\\
&\times&\sum_{n,m}{\left(P_n^{eq}+P_m^{eq}\right)\langle n|\delta\hat{A} |m\rangle \langle m|\delta\hat{B}|n\rangle\delta(E_n-E_m-\hbar\omega)}\nn
\ee
and
\be
\lefteqn{C(\omega)=2\pi \hbar}&&\\
&\times&\sum_{n,m}{P_n^{eq}\langle n|\delta\hat{A} |m\rangle \langle m|\delta\hat{B} |n\rangle\delta(E_n-E_m-\hbar\omega)}\nn
\ee
where $P_n^{eq}=e^{-E_n/k_BT}/\sum_m{e^{-E_m/k_BT}}$ is the thermal population of state $n$.
We recall the relation, 
\be
K(\omega)=2\frac{1-e^{\hbar\omega/k_BT}}{\hbar\omega/k_BT}{\rm Re}\int_0^\infty{dt e^{i\omega t} C(t)}\,. \label{K_C_relation}
\ee

\subsection{Density correlation and response function} \label{chieeSee}

We recall the well-known (fluctuation-dissipation) relation
\be 
S(\mathbf{r}_1,\mathbf{r}_2,\omega) &=& -\frac{\hbar}{2}\coth\Big(\frac{\hbar\omega}{2 k_\mathrm{B}T}\Big)\mathrm{Im}\chi(\mathbf{r}_1,\mathbf{r}_2,\omega) \label{SChifluctuationdissipation}
\ee 
between the symmetric density-density correlation function
$S(\mathbf{r}_1,\mathbf{r}_2,t)=\frac{1}{2}\langle\,\delta\hat{n}_\mathrm{e}(\mathbf{r}_1,t)\delta\hat{n}_\mathrm{e}(\mathbf{r}_2,0)+\delta\hat{n}_\mathrm{e}(\mathbf{r}_2,0) \delta\hat{n}_\mathrm{e}(\mathbf{r}_1,t)\,\rangle$
and the density-density response function
$\chi(\mathbf{r}_1,\mathbf{r}_2,t)=-\frac{i}{\hbar}\theta(t)\langle\,[\delta\hat{n}_\mathrm{e}(\mathbf{r}_1,t),\delta\hat{n}_\mathrm{e}(\mathbf{r}_2,0)]\,\rangle$.

\subsection{Lehmann representations of the density response functions} 

The Fourier transform of the density-density response function is
\begin{equation} \label{A:1}
  \begin{split}
&\chi({\bf r}_1,{\bf r}_2;\omega)\\
&=\sum_{n,m}{\frac{P_n^{eq}-P_m^{eq}}{\hbar\omega+E_n-E_m+i\eta}\langle n|\hat{n}_e({\bf r}_1)|m\rangle \langle m|\hat{n}_e({\bf r}_2)|n\rangle}\,.
  \end{split}
\end{equation}
For a system of independent particles,
\begin{equation}
  \begin{split}
&\chi({\bf r}_1,{\bf r}_2;\omega)\\
&=\sum_{n,m}{\frac{n_{FD}(\epsilon_n)-n_{FD}(\epsilon_m)}{\hbar\omega+\epsilon_n-\epsilon_m+i\eta}\langle n|\hat{n}_e({\bf r}_1)|m\rangle \langle m|\hat{n}_e({\bf r}_2)|n\rangle}\,,
  \end{split}
\end{equation}
where $n_{FD}(\epsilon) =1/\left[1+e^{-(\mu(T)-\epsilon)/k_BT}\right]$ is the Fermi-Dirac population, and $\mu(T)$ is the chemical potential.

\subsection{Lindhard response function}

The density-density response function $\chi^{0}(k,\omega)$ of a non-interacting electron gas at temeprature $T$ is given by \cite{GiulianiVignale2005}
\begin{eqnarray} \label{fullequation}
\chi^{0}(k,\omega)=-\int{\frac{d{\bf p}}{(2\pi)^{3}}\frac{n_{FD}({\bf p}+\hbar{\bf k})-n_{FD}({\bf p})}{\hbar\omega-
\epsilon({\bf p}+\hbar{\bf k})+\epsilon({\bf p})+i0+}}
\end{eqnarray}
where $\epsilon({\bf p})={\bf p}^{2}/2m$ is the energy of a particle of momentum ${\bf p}$.
Equation (\ref{fullequation}) implies \cite{DaligaultDimonte2009}
\begin{eqnarray}
\frac{\partial}{\partial\omega}{\rm Im}\chi^{0}(k,\omega=0)&=&-n\beta\sqrt{\frac{\pi m\beta}{2}}\frac{1}{k}f(k/2)\/,
\end{eqnarray}
with
\begin{eqnarray}
f(k)\equiv\frac{3\sqrt{\pi}}{4}\Theta^{3/2}n_{FD}(\hbar k)\/. \label{fofk}
\end{eqnarray}
In the classical limit ($\hbar\to 0$), 
\begin{eqnarray}
\frac{\partial}{\partial\omega}{\rm Im}\chi^{0}(k,\omega=0)&=&-n\beta\sqrt{\frac{\pi m\beta}{2}}\frac{1}{k}\/.
\end{eqnarray}

\section{Details on the derivation of the relations (\ref{Eq:5b}), (\ref{gammabargammadeltabargamma}) and (\ref{Eq:8a}).} \label{appendix_2}

Here we drop the superscript $[{\bf R}]$ indicating the dependence on the instantaneous ionic configuration, and the integral relations between the response functions are written in operator notations.
The identity operator is denoted by $\rm I$, i.e. $I({\bf r},{\bf r}')=\delta({\bf r}-{\bf r}')$.

\subsection{Relation (\ref{Eq:5b}) to the proper response} \label{appendix_2p}

Using obvious notations, the Dyson equation (\ref{Dysonproper}) can be written as
\be 
  {\chi}_\mathrm{ee}(\omega) &=& {\tilde{\chi}}(\omega) + {\tilde{\chi}}(\omega) * {v}_\mathrm{C} * {\chi}_\mathrm{ee}(\omega)\nn\\
&=& {\tilde{\chi}}(\omega) + {\chi}_\mathrm{ee}(\omega)* {v}_\mathrm{C} *{\tilde{\chi}}(\omega)  \nn\\
&=&{\varepsilon}_\mathrm{L}^{-1}(\omega) * {\tilde{\chi}}(\omega)\label{chieeepsilonLchitilde}\\
&=&{\tilde{\chi}}(\omega) * {\varepsilon}_\mathrm{R}^{-1}(\omega)\nn
\ee 
where we introduced the left and right dielectric functions
\ben 
{\varepsilon}_\mathrm{L}(\omega) = {\rm I} - {\tilde{\chi}}(\omega) * {v}_\mathrm{C}\quad,\quad {\varepsilon}_\mathrm{R}(\omega) = {\rm I} - {v}_\mathrm{C}*{\tilde{\chi}}(\omega)\,.
\een 
with inverses
\ben
{\varepsilon}_\mathrm{L}^{-1}(\omega)={\rm I}+{\chi}_\mathrm{ee}(\omega)*{v}_\mathrm{C}\quad,\quad {\varepsilon}_\mathrm{R}^{-1}(\omega)=I+{v}_\mathrm{C}*{\chi}_\mathrm{ee}(\omega)
\een
The relation ${\varepsilon}_\mathrm{L}(\omega)*{\varepsilon}_\mathrm{L}^{-1} (\omega)={\rm I}$ implies
\ben
\partial_\omega {\varepsilon}_\mathrm{L}^{-1}=-{\varepsilon}_\mathrm{L}^{-1}*\partial_\omega {\varepsilon}_\mathrm{L}* {\varepsilon}_\mathrm{L}^{-1}={\varepsilon}_\mathrm{L}^{-1}*\partial_\omega {\tilde{\chi}}(\omega) * {v}_\mathrm{C}*{\varepsilon}_\mathrm{L}^{-1}
\een
Therefore, from Eq.(\ref{chieeepsilonLchitilde}), 
\be
\partial_\omega {\chi}_\mathrm{ee}&=&\left[{\varepsilon}_\mathrm{L}^{-1}*\partial_\omega {\tilde{\chi}}* {v}_\mathrm{C}*{\varepsilon}_\mathrm{L}^{-1}\right]*{\tilde{\chi}}+{\varepsilon}_\mathrm{L}^{-1} * \partial_\omega {\tilde{\chi}}\nn\\
&=&{\varepsilon}_\mathrm{L}^{-1}*\partial_\omega {\tilde{\chi}}*\left[{v}_\mathrm{C}*{\chi}_\mathrm{ee}+{\rm I}\right] \nn\\
&=&{\varepsilon}_\mathrm{L}^{-1} * \partial_\omega{\tilde{\chi}} * {\varepsilon}_\mathrm{R}^{-1}\,. \label{Eq:A1}
\ee
By using the last expression into Eq.~(\ref{Eq:5}), we obtain the desired Eq.(\ref{Eq:5b}).

\subsection{Relations  (\ref{gammabargammadeltabargamma}) and (\ref{Eq:8a}) to the Lindhard and Kohn-Sham responses}\label{A:2}

Because of the close similarity between the Dyson equations (\ref{chieechi0}) and (\ref{DysonchichiKS}) satisfied by the free-electron and Kohn-Sham response, the derivations of Eqs.(\ref{gammabargammadeltabargamma}) and (\ref{Eq:8a}) are analogous.
We here consider the case involving the free electron response function.
the Dyson equation (\ref{gammabargammadeltabargamma}) can be written as
\ben 
{\chi}_\mathrm{ee}(\omega) &=& \chi_0(\omega) + \chi_0(\omega) * {K}(\omega) * {\chi}_\mathrm{ee}(\omega)\\
&=& \chi_0(\omega) + {\chi}_\mathrm{ee}(\omega)*{K}(\omega) * \chi_0(\omega) \\
&=&\bar{\varepsilon}_\mathrm{L}^{\,-1}(\omega) * {\tilde{\chi}}(\omega)\\
&=&{\tilde{\chi}}(\omega) * \bar{\varepsilon}_\mathrm{R}^{\,-1}(\omega)
\een 
with the frequency-dependent kernel
\ben
K(\omega)=v_{\rm C}*\left({\rm I}+G_{ee}(\omega)\right)
\een
and the left and right dielectric functions 
\ben 
\bar{\varepsilon}_\mathrm{L}(\omega) = {\rm I} - \chi_0(\omega) * K(\omega)\\
\bar{\varepsilon}_\mathrm{R}(\omega) = {\rm I} - K(\omega)*\chi_0(\omega)
\een 
with inverses
\ben
\bar{\varepsilon}_\mathrm{L}^{\,-1}(\omega)={\rm I}+{\chi}_\mathrm{ee}(\omega)*K(\omega)\\
\bar{\varepsilon}_\mathrm{R}^{\,-1}(\omega)=I+K(\omega)*{\chi}_\mathrm{ee}(\omega)
\een
Following the steps used to derive Eq.(\ref{Eq:A1}), we now obtain
\begin{equation}
  \partial_\omega{\chi}_\mathrm{ee} = \bar{\varepsilon}_\mathrm{L}^{\,-1} * \partial_\omega{\chi}_0 * \bar{\varepsilon}_\mathrm{R}^{\,-1} + {\chi}_\mathrm{ee} * \partial_\omega{G} * {\chi}_\mathrm{ee}.
\end{equation}
where the additional term results from the dependence of the kernel on the frequency.
Similarly, we find
\begin{equation}
  \partial_\omega{\chi}_\mathrm{ee} = \tilde{\varepsilon}_\mathrm{L}^{\,-1} * \partial_\omega{\chi}_{\rm KS} * \tilde{\varepsilon}_\mathrm{R}^{\,-1} + {\chi}_\mathrm{ee} * \partial_\omega{f_{xc}} * {\chi}_\mathrm{ee}.
\end{equation}
By introducing the last expressions into Eq.~(\ref{Eq:5}), we readily obtain the desired relations (\ref{gammabargammadeltabargamma}) and (\ref{Eq:8a}).

\subsection{Homogeneous limit of the response and dielectric functions} \label{appendix_homogeneous_limit}

We give properties satisfied by the reponse function $\chi_{ee}$ and related quantities in the limit of a homogeneous electron gas.
Similar relations are satisfied by $\tilde{\chi}$, $\chi_0$ and $\chi_{KS}$.

In the limit of a homogeneous electron gas,
\be
\chi_{ee}({\bf r},{\bf r}',\omega)=\chi_{ee}({\bf r}-{\bf r}',\omega)\,,
\ee
and the left and right dielectric functions are equal,
\be
\epsilon_L({\bf r},{\bf r}',\omega)=\epsilon_R({\bf r},{\bf r}',\omega)\equiv \epsilon({\bf r}-{\bf r}',\omega)\,.
\ee

The spatial Fourier transform, generally defined as
\ben
\chi_{ee}({\bf k},{\bf k}',\omega)=\frac{1}{V}\int_V{d{\bf r}e^{-i{\bf k}\cdot{\bf r}}\int_V{d{\bf r}' e^{i{\bf k}'\cdot{\bf r}'}\chi_{ee}({\bf r},{\bf r}',\omega)}}\,,
\een
satisfies
\be
\chi_{ee}({\bf k},{\bf k}',\omega)=\chi_{ee}({\bf k},\omega)\delta_{{\bf k},{\bf k}'}\,.
\ee
The inverse dielectric function satisfies
\be
\epsilon^{-1}({\bf k},\omega)=1/\epsilon({\bf k},\omega)\,.
\ee

Finally, the integral equations in Sec.~\ref{appendix_2p} and \ref{A:2} become algebraic equation, e.g.,
\be
\chi_{ee}({\bf k},\omega)&=&\tilde{\chi}({\bf k},\omega)/\epsilon ({\bf k},\omega)\\
&&\hspace{.2cm}\text{with }\epsilon({\bf k},\omega)=1-v_C({\bf k})\tilde{\chi}({\bf k},\omega) \nn\\
&=&\chi_{0}({\bf k},\omega)/\bar{\epsilon}({\bf k},\omega)\\
&&\hspace{.2cm}\text{with }\bar{\epsilon}({\bf k},\omega)=1-K ({\bf k},\omega)\chi_{0}({\bf k},\omega) \nn\\
&=&\chi_{KS}({\bf k},\omega)/\tilde{\epsilon}({\bf k},\omega)\\
&&\hspace{.2cm}\text{with }\tilde{\epsilon}({\bf k},\omega)=1-K_{KS}({\bf k},\omega)\chi_{KS}({\bf k},\omega) \nn
\ee
with $K ({\bf k},\omega)=v_C({\bf k})\left[1-G_{ee}({\bf k},\omega)\right]$ and $K_{KS}({\bf k},\omega)=v_C({\bf k})+f_{xc}({\bf k},\omega)$

\section{Sum rules} \label{appendix_3}
In this appendix we provide the proof for a set of sum
rules, the $[{\bf R}]$ superscript is dropped in order to simplify the notation.
Below,
\ben
v({\bf r})=\sum_{I=1}^{N}{v_{ie}({{\bf r}}-{\bf R}_I)}\,.
\een

\paragraph*{i)} The force matrix elements and the total linear momentum matrix elements satisfy
\begin{equation}\label{F:3B}
 \sum_{I=1}^Nf_{nm}^{Ix}=  -i \mel{n}{\hat{P}_x}{m}\frac{E_n-E_m}{\hbar}.
\end{equation}
\noindent{\it Proof:}\\
By starting from the definition
\begin{equation}\label{F:3C}
\sum_{I=1}^Nf_{nm}^{Ix} = \mel{n}{\int_V d{{\bf r}}\nabla_x v(\mathbf{r})\hat{n}_\mathrm{e}(\mathbf{r})}{m},
\end{equation}
we may rewrite the term on the right hand side as follows
\be
\int_V d{{\bf r}}\boldsymbol{\nabla} v(\mathbf{r})\hat{n}_\mathrm{e}(\mathbf{r})=\sum_{i=1}^{N_e}{\frac{\partial v(\hat{{\bf r}}_i)}{\partial\hat{{\bf r}}_i}}=-\frac{1}{i\hbar}\sum_{i=1}^{N_e}\left[{\hat{\bf p}}_i, v(\hat{\bf r}_i)\right]\nn\\
=-\frac{1}{i\hbar}\left[\sum_{i=1}^{N_e}{\hat{\bf p}}_i,
  \hat{H}_e({\bf R})\right]=-\frac{1}{i\hbar}\left[\hat{{\bf P}}, \hat{H}_e({\bf R})\right] \label{abcde}
\ee
where $\hat{{\bf P}}=\sum_{i=1}^{N_e}{\hat{\mathbf{p}}_i}$ is the linear momentum operator of the
many body system.
In deriving Eq.(\ref{abcde}), we used the relation $\left[\sum_i\hat{{\bf p}}_i,\hat{V}_{ee}\right]=0$ that results from the symmetry of the Coulomb interaction.
By substituting the previous result into Eq.~(\ref{F:3C}) we easily
obtain the final result (\ref{F:3B}).

\paragraph*{ii)} The density response function satisfies the relations:
\begin{equation}\label{F:3E}
  \int_V d{{\bf
      r}_1}\boldsymbol{\nabla}_{\mathbf{r}_1}v(\mathbf{r}_1)\chi_\mathrm{ee}(\mathbf{r}_1,\mathbf{r}_2,\omega=0)
  = \boldsymbol{\nabla}_{\mathbf{r}_2}n_\mathrm{e}(\mathbf{r}_2),
\end{equation}
\begin{equation}
  \int_V d{{\bf
      r}_2}\chi_\mathrm{ee}(\mathbf{r}_1,\mathbf{r}_2,\omega=0)
  \boldsymbol{\nabla}_{\mathbf{r}_2}v(\mathbf{r}_2) = \boldsymbol{\nabla}_{\mathbf{r}_1}n_\mathrm{e}(\mathbf{r}_1).
\end{equation}
\noindent{\it Proof:}\\
Here we limit ourself to prove the first expression, for the second one the
procedure is completely analogous. By using the Lehmann representation
for the electron-electron susceptibility (\ref{A:1})
\begin{align*}
&\int_V d{{\bf
    r}_1}\boldsymbol{\nabla}_{\mathbf{r}_1}v(\mathbf{r}_1)\chi_\mathrm{ee}(\mathbf{r}_1,\mathbf{r}_2,\omega=0)
=\\
=&\sum_{n\neq m}\frac{P_n^{\rm eq}-P_m^{\rm eq}}{E_n -
  E_m}\mel{n}{\int_V d{{\bf r}}\boldsymbol{\nabla}_{\bf r}
  v(\mathbf{r})\hat{n}_\mathrm{e}(\mathbf{r})}{m}\mel{m}{\hat{n}_{\rm
    e}({\bf r}_2)}{n}\\
=&\frac{-i}{\hbar}\sum_{n, m}(P_n^{\rm eq}-P_m^{\rm eq})\mel{n}{\hat{\bf
    P}}{m}\mel{m}{\hat{n}_{\rm e}({\bf
    r}_2)}{n}=\boldsymbol{\nabla}_{{\bf r}_2}n_{\rm e}({\bf r}_2),\\
\end{align*}
where we have used (\ref{F:3B}) for the force matrix elements, that proves Eq.~(\ref{F:3E}).

\paragraph*{iii)} 
The Kohn-Sham matrix elements satisfy
\begin{equation}\label{F:3D}
  \sum_{\mathrm{I}=1}^N \mel{n}{\delta\tilde{f}_{Ix}^{L,R}}{m}
= -i\frac{\epsilon_n-\epsilon_m}{\hbar}\mel{n}{\hat{p}_x}{m}.
\end{equation}
\noindent{\it Proof:}\\
From the definition of the force matrix elements
\ben
\sum_{{\rm I}=1}^N \mel{n}{\delta\tilde{f}_{Ix}^{L,R}}{m} =
\mel{n}{\int_V d{{\bf r}}\nabla_x v_{\rm KS}({\bf r})\hat{n}_{\rm
    e}({\bf r})}{m},
\een

we rewrite the term on the right hand side as follows
\begin{equation*}
\int_Vd{{\bf r}}\boldsymbol{\nabla}v_{\rm KS}({\bf r})\hat{n}_{\rm
  e}({\bf r}) = \frac{-1}{i\hbar}\left[\hat{{\bf p}}, \hat{V}_{\rm KS}({\bf
    r})\right] = \frac{-1}{i\hbar}\left[\hat{{\bf p}}, \hat{H}_{\rm
    KS}({\bf R})\right]
\end{equation*}
where $\hat{H}_{\rm KS}({\bf
  R})=-\frac{\hbar^2\nabla^2}{2m}+\hat{V}_{\rm KS}({\bf r})$ is the
Hamiltonian of the Kohn-Sham system, such that $\hat{H}_{\rm KS}({\bf
  R})\ket{n}=\epsilon_n\ket{n}$. From this result
(\ref{F:3D}) then follows immediately.
\paragraph*{iv)} The Kohn-Sham density response function satisfies the relations: 
\begin{equation}\label{F:3e}
  \int_V d{{\bf r}_1}\boldsymbol{\nabla}_{\mathbf{r}_1}v_{\rm
    KS}(\mathbf{r}_1)\chi_{\rm KS}(\mathbf{r}_1,\mathbf{r}_2,\omega=0)
  = \boldsymbol{\nabla}_{\mathbf{r}_2}n_\mathrm{e}(\mathbf{r}_2),
\end{equation}
\begin{equation}\label{F:3F}
  \int_V d{{\bf r}_2}\chi_{\rm KS}(\mathbf{r}_1,\mathbf{r}_2,\omega=0) \boldsymbol{\nabla}_{\mathbf{r}_2}v_{\rm KS}(\mathbf{r}_2)= \boldsymbol{\nabla}_{\mathbf{r}_1}n_\mathrm{e}(\mathbf{r}_1).
\end{equation}
\noindent{\it Proof:}\\
The procedure is analogous to the one used in $ii)$, with
the only difference that now we need to use (\ref{F:3D}) instead of
(\ref{F:3B})
\begin{align*}
  &\int_V d{{\bf r}}\boldsymbol{\nabla}_{{\bf r}_1}v_{\rm KS}({\bf
    r}_1)\chi_{\rm KS}({\bf r}_1,{\bf r}_2,\omega=0) =\\
  &=\sum_{n\neq m}\frac{p_n^{\rm eq}-p_m^{\rm eq}}{\epsilon_n -
    \epsilon_m}\mel{n}{\int_V d{{\bf r}}\boldsymbol{\nabla}_{\bf r}v_{\rm
      KS}({\bf r})\hat{n}_{\rm e}({\bf r})}{m}\mel{m}{\hat{n}_{\rm
      e}({\bf r}_2)}{n}\\
  &=\frac{-i}{\hbar}\sum_{n,m}(p_n^{\rm eq}-p_m^{\rm
    eq})\mel{n}{\hat{{\bf p}}}{m}\mel{m}{\hat{n}_{\rm e}({\bf
      r}_2)}{n}=\boldsymbol{\nabla}_{{\bf r}_2}n_{\rm e}({\bf r}_2)
\end{align*}
that proves (\ref{F:3e}), while for (\ref{F:3F}) the proof is identical.

\paragraph*{v)} The Kohn-Sham dielectric functions satisfy the relations:
\begin{equation}\label{F:3G}
  \int_V d{{\bf
      r}_1}\boldsymbol{\nabla}_{\mathbf{r}_1}v(\mathbf{r}_1)\tilde{\varepsilon}_\mathrm{L}(\mathbf{r}_1,\mathbf{r}_2,\omega=0)^{-1}
  = \boldsymbol{\nabla}_{\mathbf{r}_2}v_\mathrm{KS}(\mathbf{r}_2),
\end{equation}
\begin{equation}\label{F:3H}
  \int_V d{{\bf r}_2}\tilde{\varepsilon}_\mathrm{R}(\mathbf{r}_1,\mathbf{r}_2,\omega=0)^{-1}\boldsymbol{\nabla}_{\mathbf{r}_2}v(\mathbf{r}_2) = \boldsymbol{\nabla}_{\mathbf{r}_1}v_\mathrm{KS}(\mathbf{r}_1).
\end{equation}
\noindent{\it Proof:}\\
By using the definition of the left dielectric function
\begin{align*}
&\tilde{\varepsilon}_{\rm L}({\bf r}_1,{\bf r}_2,\omega=0)^{-1} =\\
&I({\bf r}_1,{\bf r}_2) + \int_Vd{{\bf r}_3}\chi_{\rm ee}({\bf r}_1,{\bf r}_3,\omega=0)K^{KS}({\bf r}_3,{\bf r}_2,\omega=0),
\end{align*}
into the left hand side of (\ref{F:3G}) we obtain, by using Eq. (\ref{F:3E}), (we omit the
frequency dependence in the derivation)
\begin{align*}
  &\int_V d{{\bf
      r}_1}\boldsymbol{\nabla}_{\mathbf{r}_1}v(\mathbf{r}_1)\tilde{\varepsilon}_\mathrm{L}(\mathbf{r}_1,\mathbf{r}_2,\omega=0)^{-1}
  =\\
  &=\boldsymbol{\nabla}_{{\bf r}_2}v({\bf r}_2) + \int_Vd{{\bf r}_1}\int_Vd{{\bf
      r}_3}\boldsymbol{\nabla}_{{\bf r}_1}v({\bf r}_1) \chi_{\rm
    ee}({\bf r}_1,{\bf r}_3)K^{KS}({\bf r}_3,{\bf r}_2)\\
  &=\boldsymbol{\nabla}_{{\bf r}_2}v({\bf r}_2) + \int_Vd{{\bf
      r}_3}\boldsymbol{\nabla}_{{\bf r}_3}n_{\rm e}({\bf r}_3) [v_{\rm
      C}({\bf r}_3,{\bf r}_2) + f_{\rm XC}({\bf r}_3, {\bf r}_2)]\\
  &=\boldsymbol{\nabla}_{{\bf r}_2}v({\bf r}_2) +
  \boldsymbol{\nabla}_{{\bf r}_2}v_{\rm H}({\bf r}_2) +
  \boldsymbol{\nabla}_{{\bf r}_2}v_{\rm xc}({\bf r}_2)\\
  &= \boldsymbol{\nabla}_{{\bf r}_2}v_{\rm KS}({\bf r}_2),
\end{align*}
that proves (\ref{F:3G}), while (\ref{F:3H}) may be obtained in the
same way by using the definition of the right dielectric function,
$\tilde{\varepsilon}_{\rm R}$.

\paragraph*{vi)} The friction coefficients satisfy the sum rule
\begin{equation}\label{F:3I}
  \sum_{I,J=1}^N{\gamma_{Ix,Jy}} = 0.
\end{equation}
\noindent{\it Proof:}\\
From the definition of the many body friction frictions, we
can write
\begin{align*}
  &\sum_{{\rm I,J}=1}^N{\gamma_{Ix,Jy}} =\\
  &\frac{-{\rm Im}}{M}\int_Vd{{\bf r}_1}\int_Vd{{\bf r}_2}\nabla_x
  v({\bf r}_1)\partial_\omega\chi_{\rm ee}({\bf r}_1,{\bf
    r}_2,\omega=0)\nabla_yv({\bf r}_2),
\end{align*}
while the frequency derivative of the electron-electron susceptibility is
\begin{align*}
  \partial_\omega&\chi_{\rm ee}({\bf r}_1,{\bf r}_2,\omega=0) =\\
  &-\hbar\sum_{n\neq m}\frac{P_n^{\rm eq} - P_m^{\rm eq}}{(E_n -
    E_m)^2}\mel{n}{\hat{n}_{\rm e}({\bf r}_1)}{m}\mel{m}{\hat{n}_{\rm e}({\bf r}_2)}{n},
\end{align*}
by using (\ref{F:3C}) the combination of the previous two expressions leads to
\begin{align*}
  &\sum_{{\rm I,J}=1}^N{\gamma_{Ix,Jy}} =\\
  &=\frac{\hbar{\rm Im}}{M}\sum_{n\neq m}\frac{P_{mn}^{\rm
      eq}}{E_{nm}^2}\mel{n}{\int_Vd{{\bf r}}\nabla_x\,v\hat{n}_{\rm
      e}}{m}\mel{m}{\int_Vd{{\bf r}}\nabla_y\,v\hat{n}_{\rm e}}{n}\\
  &=\frac{1}{M\hbar}{\rm Im}\sum_{n,m}(P_n^{\rm eq}-P_m^{\rm
    eq})\mel{n}{\hat{P}_x}{m}\mel{m}{\hat{P}_y}{n}\\
  &=\frac{1}{M\hbar}{\rm Im}\sum_{n,m}P_n^{\rm
    eq}\mel*{n}{\left[\hat{P}_x,\hat{P}_y\right]}{n} = 0
\end{align*}
with $P_{nm}^{\rm eq}=P_n^{\rm eq}-P_m^{\rm eq}$ and $E_{nm}=E_n-E_m$,
proving the sum rule (\ref{F:3I}).

\paragraph*{vii)} An analogous result is valid also for the Kohn-Sham
friction tensor
\begin{equation}\label{F:3L}
  \sum_{I,J=1}^N\tilde{\gamma}_{Ix,Jy} = 0.
\end{equation}
\noindent{\it Proof:}\\
From the definition of the Kohn-Sham tensor (\ref{Eq:8}) and by using the sum rules
(\ref{F:3G}) and (\ref{F:3H}) for the gradient of the external
potential it is easy to write
\begin{align*}
  &\sum_{{\rm I,J}=1}^N\tilde{\gamma}_{Ix,Jy} =\\
  &\frac{-{\rm Im}}{M}\int_Vd{{\bf r}_1}\int_Vd{{\bf r}_2}\nabla_x
  v_{\rm KS}({\bf r}_1)\partial_\omega\chi_{\rm KS}({\bf r}_1,{\bf
    r}_2,0)\nabla_y v_{\rm KS}({\bf r}_2)\\
  &\frac{\hbar{\rm Im}}{M}\sum_{n\neq m}\frac{p_{mn}^{\rm
      eq}}{\epsilon_{nm}^2}\mel{n}{\int_Vd{{\bf r}}\nabla_x v_{\rm
      KS}\hat{n}_{\rm e}}{m}\mel{m}{\int_Vd{{\bf r}}\nabla_y v_{\rm
      KS}\hat{n}_{\rm e}}{n}\\
  &=\frac{1}{M\hbar}{\rm Im}\sum_{n,m}(p_n^{\rm eq}-p_m^{\rm
    eq})\mel{n}{\hat{p}_x}{m}\mel{m}{\hat{p}_y}{n}\\
  &=\frac{1}{M\hbar}{\rm Im}\sum_{n,m}p_n^{\rm
    eq}\mel{n}{\left[\hat{p}_x,\hat{p}_y\right]}{m} = 0
\end{align*}
that finally proves (\ref{F:3L}). As a consequence of (\ref{F:3L}) and
(\ref{F:3G}) we also have
\begin{equation}
  \sum_{{\rm I,J}=1}^N\delta\tilde{\gamma}_{Ix,Jy}=0,
\end{equation}
completing the set of sum rules we seek to prove. 

\section{The electron-phonon coupling formula} \label{appendix_eph}

\subsection{Derivation of the electon-phonon coupling formula (\ref{dEdt_eph})} \label{appendix_eph_1}

We treat the general case of any monatomic Bravais lattice, whose ionic equilibrium positions are denoted by ${\bf R}^0=\{{\bf R}_I^0\}$.

The Hamiltonian of an electron gas interacting with a periodic lattice of ions oscillating around their equilibrium positions is $\hat{H}=\hat{H}_{\rm e}+\hat{H}_{\rm ph}+\hat{H}_{\rm e-ph}$, where \cite{BruusFlensberg2004}
\be
\hat{H}_{\rm e}({\bf R}^0)&=& \sum_{i=1}^{N_e}\left[\frac{{\bf p}_i^2}{2m_e}+\sum_{I=1}^{N_i}v_{ie}({\bf r}_i-{\bf R}_I^0)\right]+V_{ee}\label{He}
\ee
is the Hamiltonian of the electron gas interacting with the ions in their equilibrium positions ${\bf R}^0$,
\be
\hat{H}_{\rm ph} &=& \sum_{\lambda,{\bf q}}\hbar\omega_{{\bf q}\lambda}\bigg[\hat{b}_{{\bf q}\lambda}^\dagger \hat{b}_{{\bf q}\lambda} + \frac{1}{2}\bigg]\label{Hph}
\ee
is the Hamiltonian of the phonons, and
\be
\hat{H}_{\rm e-ph}=\frac{1}{V}\sum_{\lambda,{\bf q}}\sum_{\bf G}g_{{\bf q},{\bf G},\lambda}\hat{\rho}_{-\bfq-{\bf G}}(\hat{b}_{{\bf q}\lambda}+\hat{b}_{-{\bf q}\lambda}^\dagger) \label{Heph}
\ee
is the interaction between electrons and phonons with the phonon coupling
\be
g_{{\bf q,G},\lambda}=i\sqrt{\frac{N\hbar}{2M\omega_{\bfq,\lambda}}}({\bf q}+ {\bf G})\cdot\boldsymbol{\epsilon}_{{\bf q}\lambda}v_{ie}({\bf q}+{\bf G})\,.
\ee
Here $\hat{b}_{{\bf q}\lambda},\hat{b}_{{\bf q}\lambda}^\dagger$ are the annihilation and creation operator of a phonon of frequency $\omega_{\bfq,\lambda}$, $\boldsymbol{\epsilon}_{{\bf q}\lambda}$ are the polarization vectors, $\hat{\rho}_{\bf k}$ is the Fourier transform of the electron density, $\sum_\lambda$ is the sum over polarizabilities, $\sum_{\bfq}$ means $\bfq$ in first Brillouin zone, $\sum_{{\bf G}}$ means ${\bf G}$ in reciprocal lattice, {\bf k} is a Brillouin zone's vector, {\bf q} is localized in the first Brillouin zone and {\bf G} is a reciprocal space's vector \cite{BruusFlensberg2004}.
The interpretation of the previous expression is straightforward, the electron in fact can be scattered from any initial state $\ket{{\bf k}}$ to a final state $\ket{{\bf k}+{\bf G}+{\bf q}}$ either by absorbing a phonon in the state $\ket{{\bf q},\lambda}$ or by emitting a phonon in the state $\ket{-{\bf q},\lambda}$.

We shall apply the Fermi golden rule to calculate the rate of change of the total electron energy 
\be
\frac{dE_e}{dt} =\sum_{{\bf q},\lambda}\hbar\omega_{{\bf q}\lambda}\left[W_{\rm abs}({\bf q},\lambda)-W_{\rm em}({\bf q},\lambda)\right] \label{rate_dEdt}
\ee
where $W_{\rm abs}({\bf q},\lambda)$ is the rate of absorption and $W_{\rm em}({\bf q},\lambda)$ is the rate of emission of a phonon of energy $\hbar\omega_{{\bf q}\lambda}$ by the electronic states $\ket{m}$ defined by $\hat{H}_{\rm e}({\bf R}^0)\ket{m} = E_m\ket{m}$.
We calculate these rates  to lowest order of pertrubation theory by applying the Fermi golden rule.

\be
\hat{b}_{{\bf q}\lambda}|\dots n_{{\bf q}\lambda}\dots \rangle & = & \sqrt{n_{{\bf q}\lambda}}|\dots n_{{\bf q}\lambda}\dots \rangle
\ee
The propability per unit time of transition between state $\ket{m'}\otimes |\dots n_{{\bf q}\lambda}\dots \rangle$ and state $\ket{m}\otimes |\dots (n_{{\bf q}\lambda}-1)\dots \rangle$ is
\ben
\lefteqn{W_{\rm abs}\left(\ket{{m'}, n_{{\bf q}\lambda}}\to\ket{{m}, n_{{\bf q},\lambda} - 1}\right) } &&\\
&& = \frac{2\pi}{\hbar}\big|\mel{{m},n_{{\bf q}\lambda} - 1}{\hat{H}_{\rm e-ph}}{{m'},n_{{\bf q}\lambda}}\big|^2\delta(E_{m}-E_{m'}-\hbar\omega_{{\bf q}\lambda})\\
&& = \frac{2\pi}{\hbar}\sum_{\bf G, G'}\frac{g_{{\bf q},{\bf G},\lambda}g_{{\bf q},{\bf G'},\lambda}^*}{V^2}\mel{{m}}{\hat{\rho}_{-{\bf q}-{\bf G}}}{{m'}}\mel{{m'}}{\hat{\rho}_{{\bf q}+{\bf G'}}}{{m}}\\
&& \hspace{2.5cm}\times\, n_{{\bf q}\lambda}\delta(E_{m}-E_{m'}-\hbar\omega_{{\bf q}\lambda})\,.
\een
By averaging over a thermal distribution of electronic states at temperature $T_e$ and of phonon states at temperature $T_i$, we obtain the rate of phonon absorption
\ben
\lefteqn{W_{\rm abs}({\bf q},\lambda)} && \\
&& = \frac{2\pi}{\hbar}\sum_{m,m'}\sum_{\bf G,G'}\frac{g_{{\bf q},{\bf G},\lambda}g_{{\bf q},{\bf G'},\lambda}^*}{V^2}P_{m'}\mel{{m}}{\hat{\rho}_{-{\bf q}-{\bf G}}}{{m'}}\\
&& \times \mel{{m'}}{\hat{\rho}_{{\bf q}+{\bf G'}}}{m}N_{{\bf q}\lambda}\delta(E_m-E_{m'}-\hbar\omega_{{\bf q}\lambda}) \\
&& = \frac{1}{\hbar^2}\sum_{\bf G,G'}\frac{g_{{\bf q},{\bf G},\lambda}^*g_{{\bf q},{\bf G'},\lambda}}{V^2}C_{\bf G,G'}({\bf q},\omega_{{\bf q}\lambda})N_{{\bf q}\lambda}\,,
\een
where $N_{{\bf q}\lambda}=1/(e^{\hbar\omega_{{\bf q},\lambda}/k_BT_i}-1)$ is the Bose population of the phonon mode (${\bf q},\lambda)$ at temperature $T_i$ and $P_{m}=e^{-E_{m}/k_BT_e}/{\cal{Z}}$ is the thermal population of the electronic state $|m\rangle$.
In the second line, we have introduced the (non-symmetrical) electron density correlation function (see also appendix \ref{appendix_1})
\be
C_{\bf G,G'}({\bf q},\omega)=\int_{-\infty}^\infty dt e^{i\omega t} \langle \hat{\rho}_{{\bf q}+{\bf G}}(t)\hat{\rho}_{-{\bf q}-{\bf G'}}\rangle_e \,.
\ee

Similarly, the emission calculation gives 
\ben
\lefteqn{W_{\rm em}({\bf q},\lambda)}&&\\
&&= \frac{1}{\hbar^2}\sum_{\bf G,G'}\frac{g_{{\bf q, G},\lambda}^*g_{{\bf q, G'},\lambda}}{V^2}C_{\bf G,G'}({\bf
  q},-\omega_{{\bf q}\lambda}) (N_{{\bf q}\lambda} + 1)\\
&&= \frac{1}{\hbar^2}\sum_{\bf G,G'}\frac{g_{{\bf q, G},\lambda}^*g_{{\bf q, G'},\lambda}}{V^2}e^{-\frac{\hbar\omega_{{\bf q}\lambda}}{k_BT_e}}C_{\bf G,G'}({\bf q},\omega_{{\bf q}\lambda}) (N_{{\bf q}\lambda} + 1)\,,
\een
where in the second line we used the detailed balance property \cite{GiulianiVignale2005}.

The rate (\ref{rate_dEdt}) of energy exchange between electrons and phonons becomes
\be
\lefteqn{\frac{1}{V}\frac{dE_e}{dt}= \frac{1}{V}\sum_{{\bf q},\lambda}\sum_{\bf G,G'}\frac{\hbar\omega_{{\bf
      q}\lambda}}{\hbar^2}\frac{g_{{\bf q,
      G},\lambda}^*g_{{\bf q, G'},\lambda}}{V^2}C_{\bf G,G'}({\bf q},\omega_{{\bf
      q}\lambda})}&&\nn\\
&&\hspace{1.5cm}\times\left[N_{{\bf q}\lambda}-e^{-\hbar\omega_{{\bf
        q}\lambda}/k_BT_e}(N_{{\bf q}\lambda} + 1)\right]\label{intermediate0_dEdt}\\
&& = -2\sum_{{\bf q},\lambda}\sum_{\bf G,G'}\omega_{{\bf q}\lambda}\frac{g_{{\bf q,
      G},\lambda}^*g_{{\bf q, G'},\lambda}}{V^2}{\rm Im}\chi_{{\bf GG'}}^{[{\bf R}_0]}({\bf q},\omega_{{\bf q}\lambda})\times\nn\\
&&\hspace{1.5cm}\times\left[n_B\left(\frac{\hbar\omega_{{\bf
        q}\lambda}}{k_BT_i}\right)-n_B\left(\frac{\hbar\omega_{{\bf q}\lambda}}{k_BT_e}\right)\right] \label{intermediate_dEdt}\\
&& = \frac{4}{V}\sum_{{\bf q}}\int_0^\infty\frac{d{\omega}}{2\pi}\sum_{\bf
  G,G'} v_{ie}({{\bf q}+{\bf G}})^*v_{ie}({{\bf q}+{\bf G'}})\nn\\
&&\hspace{1.5cm}\times\,\hbar\omega {\rm Im}\chi_{{\bf GG'}}^{[{\bf R}_0]} ({\bf
  q},\omega){\rm Im}\chi_{{\bf GG'}}^{\rm ph}({\bf q},\omega) \nn\\
&&\hspace{1.5cm}\times\left[n_B\left(\frac{\hbar\omega}{k_BT_i}\right)-n_B\left(\frac{\hbar\omega}{k_BT_e}\right)\right]\,. \label{lastintermediate_dEdt}
\ee
In deriving Eq.(\ref{intermediate_dEdt}), we used the simple relation
\ben
\lefteqn{N_{{\bf q}\lambda}-e^{-\hbar\omega_{{\bf q}\lambda}/k_BT_e}(N_{{\bf q}\lambda} + 1)}&&\\
&&=\left(1-e^{-\hbar\omega_{{\bf q}\lambda}/k_BT_e}\right)\left[N_{{\bf q}\lambda}-n_B\left(\frac{\hbar\omega_{{\bf q}\lambda}}{k_BT_e}\right)\right]
\een
and the fluctuation-dissipation relation
\ben
\left(1-e^{-\hbar\omega/k_BT_e}\right)C_{\bf GG'}({\bf q},\omega)=-2\hbar V{\rm Im}\chi_{{\bf GG'}}^{[{\bf R}_0]} ({\bf q},\omega)
\een
between the correlation function and the density-density response function of electrons described by the Hamiltonian $\hat{H}_{\rm e}({\bf R}^0)$ (see Eq.(\ref{chi_eeph})).
In going from equation (\ref{intermediate_dEdt}) to the desired result (\ref{lastintermediate_dEdt}), we used the expression for the density-density response of ions in the phonon approximation derived below in Sec.~\ref{appendix_eph_chiiiph},
\be
\lefteqn{{\rm Im}\chi_{{\bf G},{\bf G'}}^{\rm ph}({\bf q},\omega) =} && \label{Imchiphphspectrum}\\
&&= -\frac{\hbar \pi n_i}{2M}\sum_\lambda \frac{1}{\omega_{{\bf q}\lambda}} ({\bf q}+{\bf G})\cdot\boldsymbol{\epsilon}_{{\bf
    q}\lambda}\,({\bf q}+{\bf G'})\cdot\boldsymbol{\epsilon}_{{\bf q}\lambda}\, \mathcal{A}_\lambda({\bf q},\omega)\nn
\ee
in terms of the phonon spectral function $\mathcal{A}_\lambda({\bf q},\omega)=\delta(\hbar\omega-\hbar\omega_{{\bf q}\lambda})-\delta(\hbar\omega+\hbar\omega_{{\bf q}\lambda})$.

Equation (\ref{lastintermediate_dEdt}) was derived by treating the electrons as a many-body system, i.e. the states $|m\rangle$ are the many-body eigenstates of the Hamiltonian $\hat{H}_e({\bf R}^0)$, Eq.(\ref{He}).
The considerations of Sec.~\ref{section_3}, where the many-body properties are expressed in terms of single-particle properties, can be straightforwardly adapted to effectively deal with electrons in a crystalline solid.
For instance, instead of using the free electron response function as in Sec.~\ref{section_3_b}, the response function $\chi_{{\bf GG'}}^{[{\bf R}_0]} ({\bf q},\omega)$ can be expressed in terms of the response function of non-interacting electrons immersed in the perfect ion lattice ${\bf R}^0$ described by the single particle Hamiltonian,
\be
\hat{H}_{\rm Bloch}({\bf R}^0)=\frac{\hat{\bf p}^2}{2m_e}+\sum_{I=1}^{N_i}v_{ie}(\hat{\bf r}-{\bf R}_I^0)\,, \label{Hbloch}
\ee
whose eigenstates are the so-called Bloch electron states.

\subsection{High temperature limit $T_{i,e}\gg\Theta_D$.} \label{appendix_eph_highT}

We show that in the hot solid limit, $T_{i,e}\gg\Theta_D$, the relation (\ref{dEdt_eph}) simplifies to Eq.(\ref{dEedtphhighT}).
The quantity ${\rm Im}\chi_{{\bf G},{\bf G'}}^{\rm  ph}({\bf q},\omega)$, which is simply related to the phonon spectrum (\ref{Imchiphphspectrum}), is non-zero only for frequency $|\omega|$ smaller than the Debye frequency $\omega_D$.
For $k_BT_{i,e}\gg\hbar\omega_D$ and $0\leq \omega\leq\omega_D$, we have
\ben
n_B(\hbar\omega/k_BT_i)-n_B(\hbar\omega/k_BT_e)\approx k_B(T_i-T_e)/\hbar\omega
\een
and, as a consequence of the small electron to ion mass ratio,
\ben
{\rm Im}\chi_{{\bf GG'}}^{[{\bf R}_0]} ({\bf q},\omega)\simeq \omega \partial_\omega {\rm Im}\chi_{{\bf GG'}}^{[{\bf R}_0]}({\bf q},0)
\een
Using these two approximations in Eq.(\ref{dEdt_eph}), we obtain 
\be
\frac{dE_e}{dt} &=&4 k_B(T_i-T_e)\sum_{{\bf q}}\sum_{\bf G,G'} v_{ie}({{\bf q}+{\bf G}})^*v_{ie}({{\bf q}+{\bf G'}})\nn\\
&&\hspace{0.5cm}\times\, \partial_\omega {\rm Im}\chi_{{\bf GG'}}^{[{\bf R}_0]} ({\bf q},0)\int_0^\infty\frac{d{\omega}}{2\pi} \omega{\rm Im}\chi_{{\bf GG'}}^{\rm ph}({\bf q},\omega)\nn\\ \label{dEedintermediatehighTlimit}
\ee
The expression (\ref{Imchiphphspectrum}) implies the relation
\ben
\int_{-\infty}^\infty\frac{d{\omega}}{2\pi} \omega{\rm Im}\chi_{{\bf GG'}}^{\rm ph}({\bf q},\omega) =-\frac{n_i}{2 M} ({\bf q}+{\bf G})\cdot({\bf q}+{\bf G'})
\een
which, when introduced in Eq.(\ref{dEedintermediatehighTlimit}), implies the desired results (\ref{dEedtphhighT}) and (\ref{geph}).

\subsection{Derivation of $g_{e-ph}$, Eq.(\ref{geph}), from the general formula (\ref{g_Kubo})} \label{appendix_eph_3}

By introducing the expression (\ref{chi_eeph}) in $g_{ei}$, Eq.(\ref{g_Kubo_b}), the average over ions disappear (it is set to ${\bf R}^0$) and we obtain
\be 
\lefteqn{g(T_e,T_i)=-\frac{k_B}{VM}{\rm Im}\sum_{\rm I=1}^N\sum_{\lambda=1}^3\frac{1}{V}\sum_{\bf q}\sum_{\bf G,G'} \partial_\omega\chi_{{\bf GG'}}^{[{\bf R}_0]} ({\bf q},0)}&&\nn\\
&\times&\int_Vd{\bf r} \nabla_\lambda v_{ie}({\bf r}-{\bf R}_I)e^{i({\bf q}+{\bf G})\cdot{\bf r}}\nn\\
&\times&\int_Vd{\bf r'} \nabla_\lambda v_{ie}({\bf r}'-{\bf R}_I) e^{-i({\bf q}+{\bf G'})\cdot{\bf r'}}\nn\\
&=&-\frac{k_Bn_i}{M}\frac{1}{V}\sum_{\bf q}\sum_{\bf G,G'} \partial_\omega {\rm Im}\chi_{{\bf GG'}}^{[{\bf R}_0]} ({\bf q},0)v_{ie}({\bf q}+{\bf G})^* \nn\\
&&\times v_{ie}({\bf q}+{\bf G}')\sum_\lambda ({\bf q}+{\bf G})\cdot\boldsymbol{\epsilon}_{{\bf q}\lambda}\,({\bf q}+{\bf G'})\cdot\boldsymbol{\epsilon}_{{\bf q}\lambda}\nn\\
&=&-\frac{k_B n_i}{M}\frac{1}{V}\sum_{{\bf q}}\sum_{\bf G,G'} v_{ie}({{\bf q}+{\bf G}})^*v_{ie}({{\bf q}+{\bf G'}})\nn\\
&&\hspace{1.cm}\times\,  ({\bf q}+{\bf G})\cdot({\bf q}+{\bf G'})\,\partial_\omega {\rm Im}\chi_{{\bf GG'}}^{[{\bf R}_0]} ({\bf q},0)\nn
\ee 

\subsection{Reduction to Allen's formula} \label{appendix_eph_allen}

In his paper, Allen describes the electrons in terms of Block states $\psi_k$, eigenstates of the Hamiltonian (\ref{Hbloch}),
\be
\hat{H}_{\rm Bloch}({\bf R}^0)\psi_k=\epsilon_k\psi_k\,,
\ee
where $k=(n,{\bf k},\sigma)$ is short for the Bloch electron quantum number (${\bf k}$ is in the first Brillouin zone, $n\in\mathbb{N}$ is the band index and $\sigma$ denotes the spin).
\be
\psi_k({\bf r})=u_{n{\bf k}}({\bf r})e^{i{\bf k}\cdot{\bf r}}\chi_\sigma\,.
\ee
The density response function of Bloch electrons satisfies
\ben
\lefteqn{{\rm Im}\chi_{{\bf GG'}}^{\rm Bloch}({\bf q},\omega)= -\frac{\pi}{V}\sum_{k,k'}\left(p_k-p_{k'}\right)}&&\\
&&\times\rho_{k,k'}({\bf q}+{\bf G}) \rho_{k',k}(-{\bf q}-{\bf G}')\delta(\hbar\omega+\epsilon_k-\epsilon_{k'})
\een
with $\rho_{k,k'}({\bf K})=\int_{V}{d{\bf r}u_{k'}^*({\bf r})u_{k}({\bf r})e^{-i({\bf k}'-{\bf k}+{\bf K})\cdot{\bf r}}}\delta_{\sigma,\sigma'}$
\ben
\lefteqn{\frac{1}{V}\frac{dE_e}{dt}}&&\\
&=& \frac{2\pi}{\hbar}\frac{1}{V} \sum_{{\bf q},\lambda}\sum_{k,k'} \hbar\omega_{{\bf q}\lambda}|M_{k,k'}^\lambda({\bf q})|^2S_{k,k'}^\lambda
\delta(\hbar\omega_{{\bf q}\lambda}+\epsilon_{k}-\epsilon_{k'})  \label{dEedt_Bloch}
\een

\ben
S_{k,k'}^\lambda&=&(p_{k}-p_{k'})\left[n_B\left(\frac{\epsilon_{k}-\epsilon_{k'}}{k_BT_i}\right)-n_B\left(\frac{\epsilon_{k}-\epsilon_{k'}}{k_BT_e}\right)\right]\\
&=&(p_{k}-p_{k'})N_{{\bf q}\lambda}+p_{k'}(1-p_{k})\,,
\een
where in the last equation we used the energy conservation described by the delta function in Eq.(\ref{dEedt_Bloch}).
where we have introduced the scattering amplitude probability
\ben 
|M_{kk'}^\lambda({\bf q})|^2 
&=&\left|\frac{1}{V}\sum_{\bf G}\rho_{k',k}(-{\bf q}-{\bf G})g_{{\bf q,G},\lambda}\right|^2\,.\\
\een 
Equation (\ref{dEedt_Bloch}) corresponds to the starting point of Allen's deriation, see equation (6) in Ref.~\cite{Allen1987}.

\subsection{The ionic density response function in the phonon approximation} \label{appendix_eph_chiiiph}

Here we derive an expression for the density-density response function of ions 
\ben
\chi({\bf k},{\bf k}',t-t') = -\frac{i}{\hbar}\frac{1}{V}\theta(t-t')\left\langle\left[\delta\hat{n}_i({\bf k},t),\delta\hat{n}_i(-{\bf k}',t')\right]\right\rangle_i
\een
in a solid at temperature $T_i$ in the phonon approximation, where $n_i({\bf k})=\sum_Ie^{-i{\bf k}\cdot{\bf R}_I}$ is the ion density and $\langle\dots\rangle_i$ denotes the thermal average at temperature $T_i$, and $\delta \hat{n}_i=\hat{n}_i-\langle \hat{n}_i\rangle_i$.
In the phonon approximation, ${\bf R}_I(t)={\bf R}_I^0+{\bf u}_I(t)$, where the harmonic displacement ${\bf u}_I$ of ion $I$ around its equilibrium position ${\bf R}_I^0$ is given by
\be
{\bf u}_I&=&\frac{1}{\sqrt{N}}\sum_{{\bf q},\lambda} \sqrt{\frac{\hbar}{2M\omega_{{\bf q},\lambda}}}(\hat{b}_{{\bf q}\lambda}+\hat{b}_{-{\bf q}\lambda}^\dagger)\boldsymbol{\epsilon}_{{\bf q}\lambda}e^{i{\bf q}\cdot{\bf R}_I^0}\nn\\
&=&\frac{1}{\sqrt{N}}\sum_{{\bf q}} {\bf u}_{\bf q}e^{i{\bf q}\cdot{\bf R}_I^0}
\ee
where ${\bf q}$ is in the first Brillouin zone.
To lowest order in the displacements,
\ben
\delta  \hat{n}_i({\bf q}+{\bf G},t) &\simeq& -i({\bf q}+{\bf G})\cdot{\bf u}_{\bf q}(t)\\
&=&-i\sqrt{\frac{\hbar}{2M\omega_{{\bf q},\lambda}}}(\hat{b}_{{\bf q}\lambda}+\hat{b}_{-{\bf q}\lambda}^\dagger)({\bf q}+{\bf G})\cdot\boldsymbol{\epsilon}_{{\bf q}\lambda}\,.
\een
This implies
\ben 
\lefteqn{\chi({\bf q}+{\bf G},{\bf q}+{\bf G}',t) \equiv\chi_{{\bf G},{\bf G'}}^{\rm ph}({\bf q},t)}&&\\
&&=\sum_\lambda\frac{\hbar n_i}{2M\omega_{{\bf q}\lambda}}D_{\rm R}^\lambda({\bf q},t)({\bf q}+{\bf G})\cdot\boldsymbol{\epsilon}_{{\bf q}\lambda}({\bf q}+{\bf G'})\cdot\boldsymbol{\epsilon}_{{\bf q}\lambda}
\een 
where $D_{\rm R}^\lambda({\bf q},t) = -\frac{i}{\hbar}\theta(t)\langle[\hat{A}_{{\bf q}\lambda}(t),\hat{A}_{{\bf q}\lambda}^\dagger(0)]\rangle$ with $\hat{A}_{{\bf q}\lambda}=\hat{b}_{{\bf q}\lambda}(t)+\hat{b}_{-{\bf q}\lambda}^\dagger$ is the retarded phonon Green's function \cite{BruusFlensberg2004}.
Using the definition of the spectral function $\mathcal{A}_\lambda({\bf q},\omega)=-{\rm Im}D_{\rm R}^\lambda({\bf q},\omega)/\pi$, we find the relation
\begin{align}
&{\rm Im}\chi_{{\bf G}.{\bf G'}}^{\rm ph}({\bf q},\omega) = && \nonumber\\
&= -\sum_\lambda\frac{\pi\hbar n_i}{2M\omega_{{\bf q}\lambda}}\mathcal{A}_\lambda({\bf q},\omega)[({\bf q}+{\bf G})\cdot\boldsymbol{\epsilon}_{{\bf q}\lambda}][({\bf q}+{\bf G'})\cdot\boldsymbol{\epsilon}_{{\bf q}\lambda}]
\end{align}
used in Sec.~\ref{appendix_eph_1}.

\section{Quick remarks about Eq.(\ref{g_binary}).} \label{appendix_phase_shifts}

The passage from Eq.(\ref{Gamma_Kubo}) to Eq.(\ref{g_binary}) in the limit of non-interacting electrons is non-trivial but can be found in several papers; e.g., see Sec. Section III-c of Ref.~\cite{dAgliano1975}.
For completeness, we recall the main steps, which rely on standard results of scattering theory.
The Kubo relation is developed as follows
\ben
\lefteqn{\Gamma=-\frac{\pi\hbar}{3M}\sum_x\int d\epsilon \frac{d n_{FD}(\epsilon)}{d \epsilon}\sum_{\bfk,\bfkp}\sum_{\sigma,\sigma^\prime}}&&\\
&&\langle \Psi_{\bfkp\sigma^\prime}^-|\hat{F}_x| \Psi_{\bfk\sigma}^+\rangle\langle \Psi_{\bfk\sigma}^+|\hat{F}_x|\Psi_{\bfkp\sigma^\prime}^-\rangle
\delta(\epsilon-\epsilon_\bfk)\delta(\epsilon-\epsilon_\bfkp)\,,
\een
over the basis of the so-called scattering states defined as
\be
|\Psi_{\bfk\sigma}^\pm\rangle=\left(1+\hat{G}^\pm\hat{t}^\pm\right)|\bfk\sigma\rangle\,.
\ee
Here $|\bfk\sigma\rangle$ is a plane wave of momentum $\hbar \bfk$, energy $\epsilon_k=(\hbar\bfk)^2/2m_e$ and spin $\sigma$; $\hat{t}^\pm=\hat{t}(\epsilon_k\pm0^+)$ and $\hat{G}^\pm=\hat{G}(\epsilon_k\pm0^+)$ with the t-matrix $\hat{t}(z)$ and resolvent operator $\hat{G}(z)=\left[z-\frac{\hat{p}^2}{2m_e}\right]$ satisfy the Lippman-Schwinger equation
\be
\hat{t}(z)=\hat{v}_{ie}+\hat{v}_{ie}\hat{G}(z)\hat{t}(z)\,. \label{Lippman_Schwinger}
\ee
We then use the property 
\ben
\langle \Psi_{\bfkp\sigma^\prime}^\pm|\hat{F}_x| \Psi_{\bfk\sigma}^\pm\rangle=i(k_x^\prime-k_x)\langle \bfkp\sigma^\prime|\hat{t}^\pm|\bfk\sigma\rangle\,,
\een
which results from $\hat{F}_x=\frac{i}{\hbar}\left[\hat{p}_x,\hat{v}_{ie}\right]$ and of properties of Eq.(\ref{Lippman_Schwinger}). 
We obtain
\ben
\lefteqn{\Gamma=\frac{\pi\hbar}{Mk_BT_e}\sum_{\bfk\bfkp}\sum_{\sigma\sigma^\prime}n_{FD}(\epsilon_\bfk)[1-n_{FD}(\epsilon_\bfk)]}&& \\
&&\hspace{1cm}\times(\bfkp-\bfk)^2|\langle \bfkp\sigma^\prime|\hat{t}^+|\bfk\sigma\rangle|^2 \delta(\epsilon_\bfk-\epsilon_\bfkp)\,.
\een
Equation (\ref{g_binary}) is then obtained from the well-known representation of the matrix elements $\langle \bfkp\sigma^\prime|\hat{t}^+|\bfk\sigma\rangle$ in terms of the phase shifts $\delta_l(k)$ for the spherically symmetric potential $v_{ie}(r)$,
\ben
\langle \bfkp\sigma^\prime|\hat{t}^+|\bfk\sigma\rangle&=&-\delta_{\sigma\sigma'}\frac{2\pi\hbar^2}{m_eVk}\\
&&\hspace{0.25cm}\times\sum_l (2l+1)e^{i\delta_l(k)}\sin\delta_l(k)P_j(\cos\Omega)
\een
with $\cos\Omega=\bfkp\cdot\bfk/k'k$.

The $T_e=0$ limit (\ref{g_binary_0}) is obtained using $dn_{FD}(\epsilon)/d\epsilon\to\delta (\epsilon-\epsilon_F)$ with $\epsilon_F=\hbar^2k_F^2/2m_e$ is the Fermi energy.

The non-degenerate limit (\ref{g_binary_nondeg}) is obtained using $n_{FD}(\epsilon)[1-n_{FD}(\epsilon)] \sim e^{(\mu-\epsilon)/k_BT_e}$ and $e^{\mu/k_BT_e}=n_e\frac{(2\pi\hbar)^3}{2}\left(2\pi m_ek_BT_e\right)^{2/3}$.

\section{Derivation of Eq.(\ref{our_Lin_et_al}).} \label{appendix_Lin_derivation}

First, we neglect the correction term $\delta\tilde{\gamma}_{\alpha\beta}^{[{\bf R}]}$ in Eq.(\ref{g_kohnsham}) and expand the expression (\ref{tildegammaab}) for $\tilde{\gamma}_{\alpha\beta}^{[{\bf R}]}$ over the Kohn-Sham states as follows
\be
\tilde{\gamma}_{\alpha\beta}^{[{\bf R}]}&=&
-\frac{\pi\hbar}{M}{\sum_{n,m}}\frac{n_{\rm FD}(\epsilon_n)-n_{\rm FD}(\epsilon_m)}{\epsilon_{n}-\epsilon_{m}}\nn\\
&&\hspace{1.cm}\times\,(\delta\tilde{f}_{\alpha}^L)_{nm}(\delta\tilde{f}_{\beta}^R)_{mn}\delta\left(\epsilon_{n}-\epsilon_{m}\right)\nn\\
&=&-\frac{\pi\hbar}{M}\iint d\epsilon d\epsilon^\prime\, \frac{p(\epsilon)-p(\epsilon^\prime)}{\epsilon-\epsilon^\prime}\delta(\epsilon-\epsilon')\nn\\
&&\times\sum_{n,m}\delta(\epsilon-\epsilon_n) \delta(\epsilon^\prime-\epsilon_{m}) (\delta\tilde{f}_{\alpha}^L)_{nm}(\delta\tilde{f}_{\beta}^R)_{mn}\,,\quad\quad\quad\label{Lin_intermediate}
\ee
where the matrix elements $f_{Ix}^{nm}=\left\langle n \big| \hat{f}_{Ix}^{(sc)}\big|m\right\rangle$ and $\hat{f}_{Ix}^{(sc)}$ is the effective force along the $x$-direction between ion $I$ and a Kohn-Sham electron screened by other electrons.
Assuming that the matrix elements depend weakly on the energies, they can be factorized outside the sum in Eq.(\ref{Lin_intermediate}), and we obtain
\ben
\tilde{\gamma}_{\alpha\beta}^{[{\bf R}]}\propto\int d\epsilon\left(-\frac{dn_{FD}(\epsilon)}{d\epsilon}\right)\left[g^{[{\bf R}]} (\epsilon)\right]^2 \,,
\een
where $g^{[{\bf R}]}(\epsilon)=\sum_n\delta(\epsilon-\epsilon_n)$ is the density of states of the Kohn-Sham system in the frozen ionic configuration ${\bf R}$.

\end{appendix}


\begin{references}

\bibitem{Gamaly2011} E.G. Gamaly, {\it Physics Reports} {\bf 508}, 91 (2011).
\bibitem{Race2010} C. P. Race, D. R. Mason, M. W. Finnis, W. M. C. Foulkes, A. P. Horsfield, and A. P. Sutton, {\it Rep. Prog. Phys.} {\bf 73}, 116501 (2010).
\bibitem{Celliers1992} P. Celliers, A. Ng, G. Xu and A. Forsman, {\it Phys. Rev. Lett.} {\bf 68}, 2305 (1992). 
\bibitem{Kaganov1957} M.I.~Kaganov, I.M.~Lifshitz and L.V.~Tanatarov, \emph{Sov. Phys. JETP} {\bf 4}, 173 (1957).
\bibitem{Fujimoto1984} J. G. Fujimoto, J. M. Liu, E. P. Ippen, and N. Bloembergen, Phys. Rev. Lett. 53, 1837 (1984).
\bibitem{Schoenlein1987} R.W. Schoenlein, W.Z. Lin, J.G. Fujimoto, and G.L. Eesley, {\it Phys. Rev. Lett.} {\bf 58}, 1680 (1987).
\bibitem{Elsayed_et_al_1987} H. E. Elsayed-Ali, T. B. Norris, M. A. Pessot, and G. A. Mourou, {\it Phys. Rev. Lett.} {\bf 58}, 1212 (1987). 
\bibitem{Allen1987} P.B. Allen, {\it Phys. Rev. Lett.} {\bf 59}, 1460 (1987). 
\bibitem{Fann1992} W. S. Fann, R. Storz, H. W. K. Tom, and J. Bokor, {\it Phys. Rev. Lett.} {\bf 68}, 2834 (1992).
\bibitem{Groeneveld1995} Rogier H. M. Groeneveld, Rudolf Sprik, and Ad Lagendijk, {\it Phys. Rev. B} 51, 11433 (1995).
\bibitem{Hohlfeld2000} J. Hohlfeld, S.-S. Wellershoff, J. G{\"u}dde, U. Conrad, V. J{\"a}hnke, and E. Matthias, {\it Chem. Phys.} 251, 237 (2000).
\bibitem{DelFatti2000} N. Del Fatti, C. Voisin, M. Achermann, S. Tzortzakis, D. Christofilos, and F. Vall{\'e}e, {\it Phys. Rev. B} {\bf 61}, 16956 (2000)
\bibitem{Bonn2000} M. Bonn, D. N. Denzler, S. Funk, M. Wolf, S.Svante Wellershoff, and J. Hohlfeld, {\it Phys. Rev. B} {\bf 61}, 1101 (2000).
\bibitem{Rethfeld2002} B.~Rethfeld, A.~Kaiser, M.~Vicanek and G.~Simon, Phys. Rev. B {\bf 65}, 214303 (2002).
\bibitem{Lin_et_al_2008} Z. Lin, L. V. Zhigilei, and V. Celli, {\it Phys. Rev. B} {\bf 77}, 075133 (2008); data available at the address http://www.faculty.virginia.edu/CompMat/electron-phonon-coupling/
\bibitem{Mueller2013} B.Y.~Mueller and B.~Rethfeld, Phys. Rev. B {\bf 87}, 035139 (2013). 
\bibitem{JiZhang2016} P. Ji and Y. Zhang, {\it Phys. Lett. A} {\bf 380}, 1551 (2016). 
\bibitem{Henighan2016} T. Henighan, M. Trigo, S. Bonetti, P. Granitzka, D. Higley, Z. Chen, M. P. Jiang, R. Kukreja, A. Gray, A. H. Reid, E. Jal, M. C. Hoffmann, M. Kozina, S. Song, M. Chollet, D. Zhu, P. F. Xu, J. Jeong, K. Carva, P. Maldonado, P. M. Oppeneer, M. G. Samant, S. S. P. Parkin, D. A. Reis, and H. A. D{\"u}rr, {\it Phys. Rev. B} {\bf 93}, 220301(R) (2016).
\bibitem{Waldecker2016} L. Waldecker, R. Bertoni, R. Ernstorfer, and J. Vorberger, {\it Phys. Rev. X} {\bf 6}, 021003 (2016). 
\bibitem{Sadasivam2017} S.~Sadasivam, M.K.Y.~Chan and P.~Darancet, Phys. Rev. Lett. {\bf 119}, 136602 (2017).
\bibitem{Maldonado2017} P. Maldonado, K. Carva, M. Flammer, and P.M. Oppeneer, {\it Phys. Rev. B} {\bf 96}, 174439 (2017).
\bibitem{Lu1} Z.~Lu, A.~Vallabhaneni, B.~Cao and X.~Ruan, {\it Phys. Rev. B} {\bf 98}, 134309 (2018).
\bibitem{Spitzer1962} L. Spitzer, Jr., {\it Physics of Fully Ionized Gases, 2nd Ed.} (Interscience, New York, 1962).
\bibitem{Ramazashvili1962} R.R. Ramazashvili, A.A. Rukhadze and V.P. Silin, {\it J. Exptl. Theoret. Phys.} {\bf 43}, 1323 (1962).
\bibitem{Kihara1963} T. Kihara and O. Aono, {\it J. Phys. Soc. Jpn.} {\bf 18}, 837 (1963).
\bibitem{Brysk1975} H. Brysk et al., {\it Plasma Phys.} {\bf 17}, 473 (1975). 
\bibitem{Hazak2001} G. Hazak, Z. Zinamon, Y. Rosenfeld, and M. W. C. Dharma-wardana, {\it Phys. Rev. E} {\bf 64}, 066411 (2001).
\bibitem{DharmaWardana_2001} M.W.C. Dharma-wardana, {\it Phys. Rev. E} {\bf 64}, 035401(R) (2001).
\bibitem{Gericke2002} D.O. Gericke, M.S. Murillo and M. Schlanges, {\it Phys. Rev. E} {\bf 65} 036418 (2002).
\bibitem{Daligault_Mozyrsky_2007} J. Daligault and D. Mozyrsky, {\it Phys. Rev. E} {\bf 75}, 026402 (2007). 
\bibitem{Daligault_Mozyrsky_2008} J. Daligault and D. Mozyrsky, {\it High Energy Density Phys.} {\bf 4}, 58 (2008). 
\bibitem{Glosli2008} J. N. Glosli, F. R. Graziani, R. M. More, M. S. Murillo, F. H. Streitz, M. P. Surh, L. X. Benedict, S. Hau-Riege, A. B. Langdon, and R. A. London, {\it Phys. Rev. E} {\bf 78}, 025401(R) (2008).
\bibitem{DimonteDaligault2008} G. Dimonte and J. Daligault, {\it Phys. Rev. Lett.} {\bf 101}, 135001 (2008) 
\bibitem{BrownSingleton2009} L.S. Brown and R.L. Singleton, {\it Phys. Rev. E} {\bf 79}, 066407 (2009). 
\bibitem{DaligaultDimonte2009} J. Daligault and G. Dimonte, {\it Phys. Rev. E} {\bf 79}, 056403 (2009). 
\bibitem{Faussurier2010} G. Faussurier, C. Blancard, P. Coss{\'e} and P. Renaudin, {\it Physics of Plasmas} {\bf 17}, 052707 (2010). 
\bibitem{Vorberger2012} J. Vorberger and D.O. Gericke, AIP Conf. Proc. {\bf 1464}, 572 (2012).
\bibitem{Baalrud2014} S.D. Baalrud and J. Daligault, {\it Physics of Plasmas} {\bf 21}, 055707 (2014).
\bibitem{Ng2012} A. Ng, {\it Int. J. Quant. Chem.} {\bf 112}, 150 (2012). 
\bibitem{Leguay2013} P. M. Leguay, A. L{\'e}vy, B. Chimier, F. Deneuville, D. Descamps, C. Fourment, C. Goyon, S. Hulin, S. Petit, O. Peyrusse, J. J. Santos, P. Combis, B. Holst, V. Recoules, P. Renaudin, L. Videau, and F. Dorchies, {\it Phys. Rev. Lett.} {\bf 111}, 245004 (2013).
\bibitem{Cho_et_al_2015} B. I. Cho, T. Ogitsu, K. Engelhorn, A. A. Correa, Y. Ping, J. W. Lee, L. J. Bae, D. Prendergast, R. W. Falcone, and P. A. Heimann, {\it Scientific Reports} {\bf 6}, 18843 (2016).
\bibitem{Jourdain_et_al_2018} N. Jourdain, L. Lecherbourg, V. Recoules, P. Renaudin, and F. Dorchies, {\it Phys. Rev. B} {\bf 97}, 075148 (2018).
\bibitem{Ogitsu2018} T. Ogitsu, A. Fernandez-Pa{\~n}ella, S. Hamel, A. A. Correa, D. Prendergast, C. D. Pemmaraju, and Y. Ping, {\it Phys. Rev. B} {\bf 97}, 214203 (2018).
\bibitem{SimoniDaligaultPRL2019} J. Simoni and J. Daligault, Phys. Rev. Lett. {\bf 122}, 205001 (2019).
\bibitem{Patterson2010} J. Patterson and B. Bailey, {\it Solid-State Physics} (Springer- Verlag, Berlin, 2010).
\bibitem{T1} \emph{Frontiers and Challenges in Warm Dense Matter}, Series: Lecture Notes in Computational Science and Engineering, {\bf 96}, edited by F.~Graziani, M.P.~Desjarlais, R.~Redmer and S.B.~Trickey (Springer 2014). 
\bibitem{Fortov2016} V.E. Fortov, {\it Extreme States of Matter, 2nd Edition} (Springer Series in Materials Science, Vol. 216, 2016). 
\bibitem{Daligault_Mozyrsky_2018} J. Daligault and D. Mozyrsky, {\it Phys. Rev. B} {\bf 98}, 205120 (2018).
\bibitem{Kubo_book} R. Kubo, M. Toda and N. Hashitsume, {\it Statistical Physics II: Nonequilibrium Statistical Mechanics} (Springer Series in Solid-State Sciences, 2003).
\bibitem{GiulianiVignale2005} G. F. Giuliani and G. Vignale, {\it Quantum Theory of the Electron Liquid} (Cambridge University Press, Cambridge, 2005).
\bibitem{Mahan2000} G.D. Mahan, {\it Many-Particle Physics, Third Edition} (Springer, 2000). 
\bibitem{BruusFlensberg2004} H. Bruus and K. Flensberg, {\it Many-Body Quantum Theory in Condensed Matter Physics: An Introduction} (Oxford University Press, 2004). In particular, see Chapters 3.6 and 16.1.
\bibitem{Kugler1975} A.A. Kugler, {\it J. Stat. Phys.} {\bf 12}, 35 (1975). 
\bibitem{SimoniDaligaultQMDpaper} J. Simoni and J. Daligault, in preparation (see also supplemental material of \cite{SimoniDaligaultPRL2019}).
\bibitem{Ulrich2012} C.A. Ulrich, {\it Time-Dependent Density-Functional Theory, Concepts and Applications} (Oxford Graduate Text, 2012).
\bibitem{Nazarov2009} V.U. Nazarov, G. Vignale and Y.-C. Chang, {\it Phys. Rev. Lett.} {\bf 102}, 113001 (2009).
\bibitem{Thiele2014} M.~Thiele and S.~K\"ummel, \emph{Phys. Rev. Lett.} {\bf 112}, 083001 (2014).
\bibitem{Wang1994} X. Y. Wang, D. M. Riffe, Y.-S. Lee, and M. C. Downer, {\it Phys. Rev. B} {\bf 50}, 8016 (1994). 
\bibitem{vanKampenbook} N.G. van Kampen, {\it Stochastic Precesses in Physics and Chemistry, 3rd Ed.} (Elsevier, 2007). See Chapter VIII.
\bibitem{ThomsonHubbard1960} W.B. Thomson and J. Hubbard, {\it Rev. Mod. Phys.} {\bf 32}, 714 (1960).
\bibitem{dAgliano1975} E. G. d'Agliano, P. Kumar, W. Schaich and H. Suhl, {\it Phys. Rev. B} {\bf 11}, 2122 (1975).
\bibitem{Dufty1996} J. W. Dufty and M. Berkovsky, p. 165 in {\it Proceedings of the International Conference on Physics of Strongly Coupled Plasmas} edited by W.D. Kraeft and M. Schlanges (World Scientific, 1996).
\bibitem{Gould1977} H. Gould and G. F. Mazenko, {\it Phys. Rev. A} {\bf 12}, 1274 (1977). 
\bibitem{Boercker1982} D. B. Boercker, F. J. Rogers, and H. E. DeWitt, {\it Phys. Rev. A} {\bf 25}, 1623 (1982).
\bibitem{FerzigerKaper1972} J.H. Ferziger and H.G. Kaper, {\it Mathematical theory of transport processes in gases} (North-Holland Publishing Company, 1972).
\bibitem{Faussurier2016} G. Faussurier and C. Blancard, {\it Phys. Rev. E} {\bf 93}, 023204 (2016).
\bibitem{StarrettSaumon2013} C. E. Starrett and D. Saumon, {\it Phys. Rev. E} {\bf 87}, 013104 (2013).
\bibitem{Starrett2017} C. E. Starrett, {\it High Energy Density Physics} {\bf 25}, 8 (2017).
\bibitem{Nazarov2005} V.U. Nazarov, J.M. Pitarke, C.S. Kim and Y. Takada, {\it Phys. Rev. B} {\bf 71}, 121106(R) (2005).
\bibitem{AshcroftMerminbook} N.W. Ashcroft and N.D. Mermin, {\it Solid State Physics} (Harcourt College Publishers, 1976); table 23.3. 
\end{references}
\end{document}